\documentclass[structabstract]{aa}

\usepackage[varg]{txfonts}
\usepackage{graphicx}
\usepackage{xcolor}
\usepackage{ulem}

\newcommand{\Kepler}{Kepler }
\newcommand{\ind}[1]{_{\mathrm{#1}}} 
\newcommand{\Dh}{D\ind{h}}
\newcommand{\Dv}{D\ind{V}}
\newcommand{\Hp}{H_P}
\newcommand{\KW}{K\ind{W}}
\newcommand{\KRM}{K\ind{RM}}

\def\d{{\rm d}}
\def\msol{M_{\odot}}

\def\tderiv#1#2{\frac{{\rm d} #1}{{\rm d} #2}}
\def\pderiv#1#2{\frac{\partial #1}{\partial #2}}

\begin{document}

\title{Seismic diagnostics for transport of angular momentum in stars}

\subtitle{ 1. Rotational splittings from the PMS to the RGB}

\author{J. P. Marques\inst{1,2}
\and
M.J. Goupil\inst{2}
\and
Y. Lebreton\inst{3,4}
\and
S. Talon\inst{5}
\and 
A. Palacios\inst{6} 
\and 
K. Belkacem\inst{2}
\and 
R.-M. Ouazzani\inst{7,2}
\and
B. Mosser\inst{2} 
\and  \\
A. Moya\inst{8}
\and
P. Morel\inst{9}
\and
B. Pichon\inst{9} 
\and
S. Mathis\inst{10,2}
\and
J.-P. Zahn\inst{11}
\and 
S. Turck-Chi\`eze\inst{10}
\and
P A. P. Nghiem\inst{10}}

\offprints{J.P. Marques}

\institute
{Georg-August-Universit\" at G\" ottingen, Institut f\" ur Astrophysik, Friedrich-Hund-Platz 1, D-37077 G\" ottingen, Germany
\and
Observatoire de Paris, LESIA, CNRS UMR 8109, F-92195 Meudon, France
\and
Observatoire de Paris, GEPI, CNRS UMR 8111, F-92195 Meudon, France
\and
Institut de Physique de Rennes, Universit\' e de Rennes 1, CNRS UMR 6251, 35042, Rennes, France
\and
D\'epartement de Physique, Universit\' e de Montr\' eal, Montr\' eal PQ H3C 3J7, Canada
\and
LUPM -  UM2/CNRS  UMR 5299 - Place Eug\` ene Bataillon cc72 - F- 34095 Montpellier, France 
\and
Institut d'Astrophysique, G\' eophysique et Oc\' eanographie de l'Universit\' e de Li\` ege, All\' ee du 6 Ao\^ ut 17, 4000 Li\` ege, Belgium
\and
Departamento de Astrof\'isica, Centro de Astrobiolog\'ia (INTA-CSIC), PO
Box 78, 28691 Villanueva de la Ca\~nada, Madrid, Spain
\and
Laboratoire Lagrange, UMR 7293, CNRS, Observatoire de la C\^ ote d'Azur,
Universit\' e de Nice Sophia-Antipolis, Nice, France.
\and
Laboratoire AIM Paris-Saclay, CEA/DSM-CNRS-Universit\' e Paris Diderot, IRFU/SAp Centre de Saclay, 91191 Gif-sur-Yvette, France
\and
Observatoire de Paris, LUTH, CNRS UMR 8102, F-92195 Meudon, France}

\date{\ today }

\abstract
{Rotational splittings are currently measured for several main sequence stars and
a large number of red giants with 
the  space mission Kepler. This will provide stringent constraints on rotation profiles.}
{ Our aim is to obtain 
seismic constraints on the internal transport and
surface loss of angular momentum  of oscillating solar-like stars. 
To this end, we study the evolution of 
 rotational splittings  from the pre-main sequence to the red-giant
branch for stochastically excited oscillation modes. }
{We  modified the evolutionary code CESAM2K  to take rotationally induced
  transport  in radiative zones into account. Linear rotational splittings were computed  for a
 sequence of $1.3 M_{\odot}$ models.
 Rotation profiles were derived from our evolutionary  models 
and eigenfunctions from linear adiabatic oscillation calculations.} 
{ We  find that transport by meridional circulation and shear turbulence 
 yields  far too high a core rotation rate for red-giant models compared with recent
seismic observations.  We discuss several 
uncertainties in the physical description of stars 
 that could have an impact on the
rotation profiles.
 For instance, we find that the
 Goldreich-Schubert-Fricke instability does not extract
 enough angular momentum from the core to account for the discrepancy. 
In contrast,  an increase of the 
  horizontal  turbulent viscosity by 2 orders of magnitude  
  is able to significantly decrease the central rotation rate on
  the red-giant branch.
 }
{Our results  indicate that 
 it is possible that the prescription for the 
 horizontal turbulent viscosity  largely underestimates its actual value or else
a mechanism not included  in current stellar models of
low mass stars is needed to slow down the rotation in the radiative core
 of red-giant stars. }


\maketitle

\section{Introduction} %

Stars rotate, and this rotation has important consequences on their evolution.  
On the one hand, centrifugal acceleration reduces local gravity, mimicking a lower mass. 
On the other hand, rotation induces meridional circulation \citep{1925Obs....48...73E,1953MNRAS.113..716M} and
shear and baroclinic instabilities \citep{2004A&A...425..243M}, which contribute to the mixing of chemical elements.

The problem of transport of angular momentum inside stars has not yet been fully understood. Several mechanisms seem to be active,
 the most  commonly invoked being diffusion by turbulent viscosity, transport by meridional circulation \citep{1992A&A...265..115Z, 
 1998A&A...334.1000M}, torques due to magnetic  fields \citep{2004A&A...422..225M,2005A&A...440..653M,2011A&A...532A..34S},
 and transport by gravity waves \citep{2005A&A...440..981T}.
  For stars with significant convective envelopes ($M_{\star} \lesssim 1.4 M_{\odot}$), the problem is complicated further by the 
   magnetic braking by stellar winds \citep{1988ApJ...333..236K}. Another problem concerns the initial rotation state, 
  which can have an influence on the rotation profile during the pre-main sequence (PMS) and early main sequence (MS). 
The initial rotation state depends on the presence 
  and lifetime of a circumstellar disk during the first stages of the PMS; magnetic fields seem to effectively lock the stellar surface 
  to the disk \citep{1994ApJ...429..781S}. \citeauthor{2009pfer.book.....M}'s \citeyear{2009pfer.book.....M} textbook contains an 
extensive description of these processes.


Several stellar evolutionary codes already include transport of angular momentum and associated chemical element mixing 
 \citep{1995ApJ...441..865C, 1997A&A...322..209T, 2003A&A...399..603P, 2008Ap&SS.316...43E}. 
 They have been developed first  with a focus on studying  consequences on the evolution of 
 massive stars  or evolved stars \citep[see, e. g.,][]{2000ARA&A..38..143M,2000ApJ...540..489S}. The Sun and 
solar-like stars have been studied as above
\citep[e. g.,][]{1995ApJ...441..865C, 1997A&A...322..320Z, 1997PhDT........18T}, and one important application was the lithium surface depletion 
\citep[e. g.,][]{1998A&A...335..959T, 1999A&A...351..635C}.

One way to study the internal transport and  evolution of angular momentum
is to obtain seismic information
on the internal rotation profile of stars in the low and intermediate mass
range at different stages of evolution. The SoHO satellite for the Sun \citep{2010ApJ...715.1539T} and 
the  ultra high precision photometry (UHP) asteroseismic space missions 
 CoRoT \citep{Baglin:2003} and \Kepler \citep{2010Sci...327..977B} 
offer such an opportunity. 
Seismic information has being obtained for 
PMS and  MS stars \citep[e. g.,][]{2010AN....331..956B, 2012A&A...543A..96E}, and 
now we also have  seismic observational constraints
on the internal rotation of red-giant stars
\citep[e. g.,][]{2012Natur.481...55B, 2012arXiv1209.3336M, 2012ApJ...756...19D}.

Our goal   is to use seismic diagnostics  to  test the   description of 
transport of angular momentum  processes in 1-D stellar models of  
low-mass stars,  from the PMS to the red-giant branch (RGB).
Rotational splittings
(differences between nonaxisymmetric oscillation modes) are such diagnostics.
 When the rotation is slow, the relation between rotational splittings and rotation rate is linear and
therefore relatively easy to interpret. 

 The information that rotational splittings  provide on the internal rotation,
   however, depends on the physical nature of the stochastically excited  modes,  which in turn
   depend on the structure of the star, hence its age. In addition, some rotation gradient is expected to develop
during the evolution; 
the core rotation rate is expected to become higher than the surface, as the core contracts and the envelope expands.
If the rotation rate of  some layers becomes too high,  the associated distortion of the star 
and/or the Coriolis acceleration invalidates the linear approximation that provides
 the usual (linear) 
rotational splittings. In such cases, nonperturbative methods  must be used
\citep[such as those developed in ][]{2006A&A...455..607L,2006A&A...455..621R,Ouazzani2012}.
It is therefore necessary to determine if linear splitting are valid for an evolutionary
stage in order to interpret correctly the observations.

We have thus started a series of papers devoted to establish seismically 
 validated  processes of transport of angular momentum that play an essential role in shaping the rotation profile of 
  stars. 
The present paper  is the first of this  series. The second and the third  will focus
specifically on  the case of red-giant stars. They
will investigate the seismic diagnostics of rotation 
 for slowly rotating red-giant stars for which linear rotational splittings are valid, and
 then rapidly rotating red-giant stars with nonperturbative methods.

  In the present paper,  we compute the rotation profiles and 
  their evolution with time from the PMS to the RGB
 with an evolutionary code 
where  rotationally induced mixing  using the
prescription of \cite{1992A&A...265..115Z} as refined by \cite{1998A&A...334.1000M} has been implemented. We then
follow the evolution of the  linear rotational splittings 
 calculated with the rotation
 profiles obtained from our evolutionary models and discuss their validity.

The paper is organized as follows: in Section \ref{sec:physinps} we describe 
the physical inputs that are implemented in our evolutionary code, with special emphasis
on the transport of angular
momentum   and rotation induced mixing.  
 We developed a version CESAM2K of the  code CESAM \citep{Morel:1997, 2008Ap&SS.316...61M}
 by implementing rotation-induced transport
  in radiative zones with careful attention to conservation
  and transport of angular momentum. 
  The numerical scheme for the
 rotationally induced transport had to be modified compared to the general scheme used in CESAM2K,
and  some differences in the physical inputs also exist.
For these reasons, the modified version will be referred to as the
  CESTAM code
 hereafter. This will avoid any possible confusion with 
  results from other
 versions of CESAM2K used in the international community. The  acronym 
CESAM stands for Code d'Evolution Stellaire 
Adaptatif et Modulaire, and the extra {\it T} in CESTAM stands for transport.
A  validated   standard version of CESTAM  
 will be made freely available.

The numerical implementation is described in Section \ref{sec:numerical}.
 We then show comparisons with other codes for validation 
(Section \ref{sec:compare}).  During the implementation of the code, we encountered some 
difficulties already present but not clearly explained in previous work. We chose to describe them 
carefully. 
 In Sect. \ref{sec:evsplit} we  follow the evolution 
of  theoretical  linear rotational
splittings for  a 1.3 $M_\odot$ sequence of stellar models evolved from the PMS to the RGB. 
 The $\ell$ $=1,2$ modes are computed  with  the ADIPLS code \citep{2008Ap&SS.316..113C} 
in a frequency 
range where the modes are expected to be
stochastically excited. 
We discuss the behavior of the splittings with frequency
 that can  be observed during the various phases of evolution. For red-giant models, we find rotational
splittings that are much higher than observed. 
 In Section \ref{sec:discussion}    we consider  possible reasons for this disagreement. 
Finally, conclusions and some perspectives are given in Sect. \ref{sec:conclusions}.

\section{Physical input}\label{sec:physinps}%

The original version of the CESAM2K code \citep{Morel:1997, 2008Ap&SS.316...61M}  consists of a set of
 routines that calculates 1-D quasi-hydrostatic stellar evolution 
including microscopic diffusion of chemical species. The solution of the quasi-static equilibrium is performed by a collocation 
method based on piecewise polynomial approximations projected on a B-spline basis. 
For models without diffusion, the evolution of the chemical composition is solved by stiffly stable schemes of orders up to four; in the 
convection zones mixing and evolution of chemicals are simultaneous. The solution of the diffusion equation employs the Galerkin 
finite elements scheme, and the mixing of chemicals in convective zones is then performed by a strong turbulent diffusion.

The code CESAM2K allows the choice of several options for the physics. The microscopic input physics is updated regularly. The opacity 
tables are presently the OPAL95 data \citep{Iglesias/Rogers:1996} complemented at low temperatures by the Wichita opacity data 
\citep{2005ApJ...623..585F}.  Several sets of opacity tables are provided that correspond to various mixtures of chemical elements, 
for instance the \citet{1993PhST...47..133G} or \citet{2005ASPC..336...25A,2009ARA&A..47..481A} solar mixtures,
 as well as $\alpha$-element enhanced 
mixtures. We included Pothekhin's updated conductive opacities \citep[see e. g.][]{2007ApJ...661.1094C}.
 Several options are possible for the equation of state, the most commonly used being the OPAL2005 EoS 
\citep{Rogersetal:1996, Rogers/Nayfonov:2002}. Several networks of nuclear reactions (and corresponding updated nuclear 
reaction rates) were implemented, allowing the evolution of stars to be calculated from the PMS up to helium burning. The
 microscopic diffusion of chemical elements includes gravitational settling, thermal, and concentration diffusion terms, but 
no radiative accelerations. Two formalisms are available for microscopic diffusion transport 
 \citep{Burgers:1969, Michaud/Proffitt:1993}. Also, two options are available for treating convection, the classical 
MLT theory of \citet{Bohm-Vitense:1958} or the \citet{Canutoetal:1996} so-called full spectrum of turbulence. The atmospheric 
boundary condition is derived either from gray model atmospheres or from Kurucz ATLAS9 models \citep{2005MSAIS...8...14K}. 
Overshooting is an option.

Mass loss can be considered with different prescriptions. In the models presented hereafter we use 
the empirical mass loss rates of \citet{1975MSRSL...8..369R} scaled with metallicity according to \citet{1992A&AS...96..269S}.

Recently, the CESAM2K code has been involved in the ESTA activities undertaken to prepare the interpretation of the seismic 
observations of CoRoT. Stellar models were calculated for a range of mass, chemical composition, and the evolutionary stages 
corresponding to CoRoT main targets and have been compared with the results of several other evolutionary codes showing a very
 good general agreement \citep{2008Ap&SS.316..187L, 2008Ap&SS.316....1L, 2008Ap&SS.316..219M}.

 CESAM2K is freely available for download, with the details in \citet{2008Ap&SS.316...61M}. 

\subsection{Transport of angular momentum}

In convective zones, although there is differential rotation in latitude, the mean rotation rate at a given radius weakly depends 
 on the radius. Therefore,  it is often assumed in 1-D stellar evolution codes that convective zones rotate as solid bodies 
\citep[e. g.][]{1997A&A...322..209T, Meynet/Maeder:2000, 2003A&A...399..603P}.  In extended convective zones, however, comparisons with 3-D
numerical simulations suggest instead a prescription of uniform specific angular momentum 
\citep[e. g.][]{2000MNRAS.316..395D, 2006A&A...453..261P}. 
Both options can be used in CESTAM.

In a radiative zone, 
we used the formalism of  \citet{1992A&A...265..115Z}, refined in  \citet{1998A&A...334.1000M}, to model the transport of angular 
momentum and chemical species by meridional circulation and shear-induced turbulence. In what follows, we sketch the model to
 make it clear exactly which equations we adopted, as several slightly different versions have been used 
\citep[e. g.][]{1992A&A...265..115Z, 1997A&A...322..209T, 1998A&A...334.1000M, 2000MNRAS.316..395D, 2003A&A...399..603P, 2004A&A...425..229M}.

Turbulence is expected to be highly anisotropic owing to the stable 
stratification in radiative zones. Turbulence would then be much stronger 
 in the horizontal than in the vertical direction. Thus, the hypothesis of 
 ``shellular rotation'' can be used: as differential 
 rotation is presumably weak along isobars, it is treated as a perturbation. 
 All variables $f$ can be split into a mean value over 
 an isobar and a perturbation \citep[as in][] {1992A&A...265..115Z}:
\begin{equation}
 f\left(p,\vartheta\right) = \overline{f}(p) + \tilde{f}_2(p) P_2(\cos \vartheta),
\end{equation}
where $P_2(\cos \vartheta)$  is the second-order Legendre polynomial and $p$ is the pressure.
Higher order effects, included 
in the formalism developed by \cite{2004A&A...425..229M}, are not considered here for the moment.

The velocity of meridional circulation can be written 
in a spherical coordinate system as
\begin{equation}
  \overrightarrow{U} = U_2(r) P_2(\cos \vartheta) \overrightarrow{e}_r + V_2(r) 
\frac{\d P_2(\cos \vartheta)}{\d \vartheta} \overrightarrow{e}_{\vartheta}, \label{eq:uvec}
\end{equation}
where $r$ is the mean radius of the isobar and $\vartheta$ the colatitude. The vertical component $U_2$ 
is given in Appendix \ref{sec:detail}, following \citet {1998A&A...334.1000M}.
The equation of continuity in the anelastic approximation gives then the horizontal component $V_2$:
\begin{equation}
 V_2 = \frac{1}{6 \rho r} \frac{\d}{\d r}\left(\rho r^2 U_2 \right),
\end{equation}
where $\rho$ is the density.

The transport of angular momentum obeys an advection-diffusion equation:
\begin{equation}
 \rho \frac{\d}{\d t} \left(r^2 \Omega \right) = \frac{1}{5 r^2} \frac{\partial}{\partial r} 
\left(\rho r^4 \Omega U_2 \right) + \frac{1}{r^2} \frac{\partial}{\partial r} \left(r^ 4 \rho \nu\ind V 
\frac{\partial \Omega}{\partial r} \right),  \label{eq:transp}
\end{equation}
where $\nu\ind V$ is the vertical component of the turbulent viscosity and $\d/\d t$ represents the Lagrangian time derivative.

The relative horizontal variation of the density, $\Theta = \tilde{\rho}/\rho$, and the mean molecular weight, 
$\Lambda = \tilde{\mu}/\mu$, obey
\begin{eqnarray}
 \Theta &=& \frac{2 r^2}{3 g} \Omega \frac{\partial \Omega}{\partial r} \label{eq:theta}\\
 \frac{\d \Lambda}{\d t} &=& U_2 \frac{\nabla_{\mu}}{\Hp} - \frac{6 \Dh}{r^2} \Lambda, \label{eq:lambda}
\end{eqnarray}
where $\Dh$ is the horizontal component of turbulent diffusion discussed in Appendix \ref{sec:visc}. The evolution of $\Lambda$,
 Eq. (\ref{eq:lambda}), depends on the competition between the advection of a mean molecular weight gradient 
and its destruction by the horizontal turbulent diffusion.

\subsection{Evolution of the chemical composition}

The vertical advection of chemicals due to the large-scale meridional 
circulation coupled with a strong horizontal turbulent diffusion results 
in a vertical diffusion process \citep[see, e.~g.][]{1992A&A...253..173C}. 
The equation of the chemical composition evolution can then be written as
\begin{equation}
 \tderiv{X_i}{t} = \frac{\partial}{\partial m} \left[ \left(4 \pi r^2 \rho \right)^2 \left(\Dv + D\ind{eff} \right) 
\pderiv{X_i}{m} \right] + \left. \tderiv{X_i}{t} \right|_{\rm nucl} + \left. \tderiv{X_i}{t} \right|_{\rm micro}
\end{equation}
where $X_i$ is the abundance by mass of the $i$-th nuclear species and 
$\Dv=\nu\ind V$ 
and $D\ind{eff}$ the vertical diffusivity and the diffusion coefficient associated with the meridional circulation; $\Dv$ 
and $D\ind{eff}$ are discussed in Appendix \ref{sec:visc}.

A necessary condition for shear instability is the Richardson
 criterion as given by \citet{1997A&A...317..749T}.  Another condition is that
 the turbulent viscosity $\nu\ind V$ must be
   greater than the molecular viscosity, $\nu$,
   as expressed by the Reynolds criterion 
\begin{equation}
 \nu_V > \nu \, {\rm Re}\ind c, \label{eq:reyn},
\end{equation}
where
${\rm Re}\ind c \simeq 10$ is the critical Reynolds number \citep{2000A&A...364..876S}.
 When condition (\ref{eq:reyn}) is not satisfied we use $\Dv = \nu\ind{V} = \nu$.

\subsection{Initial conditions}

Stars are fully convective when they start their PMS evolution on the Hayashi track. Assuming that 
convective zones rotate like solid bodies, the star should have uniform angular velocity at the beginning of its evolution. 

Several facts complicate this simple picture. First, \cite{Palla/Stahler:1991} showed that stars that are more
 massive than about $2 M_{\odot}$ are no longer fully convective when they appear on the PMS (the exact 
 mass depends mainly on the protostellar accretion rate and deuterium abundance). Second, if during the PMS the 
only process slowing down stellar rotation were the magnetic braking by stellar winds mentioned above, stars would 
rotate much more rapidly than observed on the ZAMS. The PMS is too short for this process to slow down the star significantly.
 An additional process is needed during the PMS, most likely disk locking \citep[e. g.,][] {1997A&A...326.1023B}.

Young stars are most often surrounded by a circumstellar disk left over after the main accretion phase is over.
 The magnetic coupling between the star and the disk slows the star down
 \citep[see, e. g.,][]{1994ApJ...429..781S}, and this effect is often modeled by assuming that 
 the stellar surface corotates with the disk at a constant angular velocity
  \citep[see][]{1997A&A...326.1023B}  as long as the disk exists. 
  Once the disk has disappeared, the star surface rotation evolves freely. 
 This scenario is implemented in CESTAM  with 
  the disk lifetime $\tau_{\rm disk}$ and period $P_{\rm disk}$ as free parameters of the model.

\subsection{Magnetic braking}

Stars that are less massive than about $1.4 M_{\odot}$ have significant outer convective zones.
 A solar-type dynamo operates there and generates a magnetic field. The coupling between the
  magnetic field and the plasma in the stellar wind strongly brakes the rotation of the star. 
 \citet[hereafter K88]{1988ApJ...333..236K} proposed the following law for the loss of angular momentum $ \dot{J}$:
\begin{eqnarray}
 \dot{J} &=& -\KW \Omega^3 \left(\frac{R}{R_{\odot}} \right)^{\frac 1 2}  
\left(\frac{M}{M_{\odot}} \right)^{-\frac 1 2} \,\, {\rm for} \; \Omega < \Omega_{\rm sat} \nonumber \\
 \dot{J} &=& -\KW \Omega\, \Omega_{\rm sat}^2 \left(\frac{R}{R_{\odot}} \right)^{\frac 1 2}  
\left(\frac{M}{M_{\odot}} \right)^{-\frac 1 2} \,\, {\rm for} \; \Omega \geq \Omega_{\rm sat},\label{eq:kawaler}
\end{eqnarray}
where $\Omega_{\rm sat}$ is a saturation angular velocity, above which magnetic field generation seems to saturate.
The value of $\Omega_{\rm sat}$ is often set at $8-14\Omega_{\odot}$ \citep[as in][]{1997A&A...326.1023B}. 
There are indications, however, that $\Omega_{\rm sat}$ varies with stellar mass. 
\citep[see, e. g.][]{1997ApJ...480..303K, 2003ApJ...582..358A}.

 The parameter $\KW$ in Eq. (\ref{eq:kawaler}) is usually calibrated by requiring that 
calibrated solar models have $\Omega=\Omega_{\odot}=2.86 \times 10^{-6}\,{\rm rad}\,{\rm s}^{-1}$. The precise 
value of $\KW$ needed to spin down the Sun to its current period depends on the prescription for the transport 
of angular momentum adopted (see section \ref{sec:sun} below).
The parameter  $\KW$ should also depend on stellar mass \citep[see, e. g.][]{1997ApJ...480..303K}.

Recently, \citet[hereafter RM12]{2012ApJ...746...43R} have criticized the approach used in K88. Specifically, 
K88 supposed that the surface magnetic flux goes as some power of the angular velocity, whereas
RM12 suggest instead that it is the magnetic field strength that obeys such a law. As a consequence, they obtain
the following law:

\begin{eqnarray}
 \dot{J} &=& -\KRM \Omega^5 \left(\frac{R}{R_{\odot}} \right)^{\frac 8 3}  
\left(\frac{M}{M_{\odot}} \right)^{-\frac 2 3} \,\, {\rm for} \; \Omega < \Omega\ind{sat} \nonumber \\
 \dot{J} &=& -\KRM \Omega\, \Omega\ind{sat}^4 \left(\frac{R}{R_{\odot}} \right)^{\frac 8 3}  
\left(\frac{M}{M_{\odot}} \right)^{-\frac 2 3} \,\, {\rm for} \; \Omega \geq \Omega\ind{sat}.\label{eq:reiners}
\end{eqnarray}
The value of $\KRM$ does not depend on the stellar mass, and $\Omega\ind{sat} \simeq 3 \Omega_{\odot}$. Both 
prescriptions, Eqs.~(\ref{eq:kawaler}) and (\ref{eq:reiners}), are implemented in CESTAM and tested below.

\section{Numerical procedure}\label{sec:numerical}%

\subsection{The overall problem}

Stellar evolution depends on two interconnected problems: the problem of stellar structure 
(solving the stellar structure equations), and the problem of the evolution of the chemical composition. 
It has proved very difficult to solve the two problems simultaneously; 
different stellar evolution codes employ several techniques to overcome this difficulty. Some codes
 \citep[e. g.,][]{2008Ap&SS.316...25D} compute the solutions of the structure equations and chemical composition 
in an independent way, where they use the structure calculated at the previous time step to evolve the chemical composition to the 
current time step, and then use this composition to calculate the structure at the current time step (or the other 
way around: first the structure, then the chemical composition). A second kind of code computes the chemical composition 
between each iteration of the structure problem, as in \citet{2008Ap&SS.316...83S}. And finally, Eggleton's code
 \citep{1971MNRAS.151..351E} solves the two problems simultaneously. \citet{2006MNRAS.370.1817S} has shown that differences 
between the three kinds of codes are only significant on the AGB.

The CESAM2K code \citep{Morel:1997, 2008Ap&SS.316...61M} belongs to 
the second category above. It
 begins a time step by updating 
the chemical composition. With the new chemical composition, a new structure is obtained 
by performing one iteration of the algorithm used to solve the structure equations. This structure 
is used to update the chemical composition again, before a new iteration on the structure is performed. 
The procedure is repeated until convergence of the structure algorithm. CESTAM keeps this iterative scheme.

Rotation with angular momentum transport introduces a new problem interconnected with 
the previous two. The evolution of the chemical composition depends on the turbulent mixing 
induced by differential rotation, whereas the structure equations are changed by the inclusion
 of the centrifugal acceleration. In CESTAM, we inserted the resolution of the angular momentum transport 
  equations at the beginning of the cycle, because the rotation profile is required to calculate 
  the turbulent diffusion coefficients that are needed to update the chemical composition. 
  Our cycle, then, is as follows: we update the rotation profile, then the chemical composition,
   and finally we iterate on the structure. The cycle is repeated until convergence.

In the absence of external torques, total angular momentum is conserved. The rotation profile obtained at this stage does 
not enforce angular momentum conservation, however, because the profile $\Omega(m)$ was computed with the stellar structure obtained 
in the iteration before convergence. The total angular momentum is $J \propto \int_0^m \Omega(m) r^2(m) \d m$ and the 
function $r(m)$ changes between iterations. We need to compute the rotation profile one more time after convergence 
to make sure that it is consistent with the stellar structure, so that angular momentum is numerically conserved.

\subsection{The rotation profile}

Equation (\ref{eq:transp}) (with Eq. (\ref{eq:u2}), Eq. (\ref{eq:theta}) and Eq. (\ref{eq:lambda})), is a
 fourth-order differential equation in $r$. To solve it, we split it into four first-order differential 
 equations, and solve the system using the well known relaxation method \citep{Henyeyatal:1964, numrecipes}.
  Equation (\ref{eq:lambda}) is solved simultaneously with Eq. (\ref{eq:transp}), so that we have a total of five
  finite difference equations. We chose this method for its simplicity in dealing with the complex relations between 
variables expressed in Eq. (\ref{eq:transp}), taking Eqs. (\ref{eq:u2}), (\ref{eq:theta}), and (\ref{eq:lambda}) into account. 
Methods that require that the solutions are approximated by a linear combination of known functions (collocation methods, 
spectral methods) are therefore difficult to implement. In practice, the relaxation method we employed proved to be efficient, 
robust, and fairly stable.

The scheme can be fully implicit or semi-implicit in time (to ensure higher order accuracy in time). The time step can be 
subdivided if needed, a useful feature for future developments involving faster processes.

Four boundary conditions are needed. At the top of a radiative zone, we impose conservation of 
angular momentum and no differential rotation: 
      \begin{eqnarray}
       && \frac{\partial \Omega}{\partial r} = 0 \nonumber \\
	&& \frac{\d}{\d t} \int_{m_t}^{M_{\star}} \left(r^2 \Omega \right) \d m = 
-\left. \frac {4 \pi} 5 \rho r^4 \Omega U_2 \right|_{m=m_t} + \frac 3 2 \dot{J},
      \end{eqnarray}
where $m_t$ is the mass inside the top of the radiative zone. If the convective zone above it is at the surface, $\dot{J}$ is 
the torque applied at the surface of the star, Eq. (\ref{eq:kawaler}), otherwise $\dot{J}=0$. At the bottom of a radiative zone, similarly, 
      \begin{eqnarray}
       && \frac{\partial \Omega}{\partial r} = 0 \nonumber \\
	&& \frac{\d}{\d t} \int_{0}^{m_b} \left(r^2 \Omega \right) \d m = \left. \frac {4 \pi} 5 \rho r^4 \Omega U_2 \right|_{m=m_b},
 \label{eq:boundb}
      \end{eqnarray}
where $m_b$ is the mass inside the bottom of the radiative zone. If the center of the star is radiative ($m_b=0$), Eq. 
(\ref{eq:boundb}) is replaced by  $U_2=0$, the requirement that there is no mass flow out of the center.

Intermediate convective regions (regions that are neither at the center nor at the surface) are treated 
as if they were special points. 
The equations between the beginning and the end of an intermediate convective zone (between $m\ind i$ and $m\ind f$) are
\begin{eqnarray}
 \frac{\d}{\d t} \int_{m\ind i}^{m\ind f} \left(r^2 \Omega \right) \d m &=& 
\frac{4 \pi} 5 \left(\rho\ind i r\ind{i}^4 \Omega\ind i U\ind{2,i} - \rho\ind f r\ind{f}^4 \Omega\ind f U\ind{2,f}  \right) \label{eq:conv1}\\
  \left. \frac{\partial \Omega}{\partial r}\right|\ind i &=& 0 \label{eq:conv2} \\
  \left. \frac{\partial \Omega}{\partial r}\right|\ind f &=& 0 \label{eq:conv3} \\
 \Lambda\ind i &=& 0 \label{eq:conv4} \\
 \Lambda\ind f &=& 0 \label{eq:conv5}.
\end{eqnarray}
Equation (\ref{eq:conv1}) guarantees conservation of total angular momentum, Eqs. (\ref{eq:conv2}-\ref{eq:conv3}) impose no
 shear at the borders of the convective zone, and Eqs. (\ref{eq:conv4}-\ref{eq:conv5}) result from the absence of 
$\mu$-gradients in a convective zone.

The four equations resulting from Eq. (\ref{eq:transp}) are written in finite-difference form between 
pairs of points, say, at $m=m_{k-1}$ and $m=m_k$. Equation (\ref{eq:lambda}) requires a special 
treatment, because it is not a differential equation in 
$r$ (or $m$). We write Eq. 
(\ref{eq:lambda}) as an algebraic equation at point $m=m_{k-1}$, because otherwise we would be solving for 
$\left(\Lambda_k + \Lambda_{k-1} \right)/2$ (the average between $m_{k-1}$ and $m_k$ instead of the values at 
$m_{k-1}$ and $m_k$). Indeed, $\Lambda$ only appears as $\left(\Lambda_k + \Lambda_{k-1} \right)/2$ in the other 
equations, since it has  no space derivatives. A fifth boundary condition is needed: $\Lambda=0$ at $m = m_t$ 
(the top of the radiative zone) as in Eqs. (\ref{eq:conv4}-\ref{eq:conv5}).

\begin{table}
 \caption{Parameters of calibrated solar models.
   } 
 \label{tab:sunparams} 
 \centering 
 \begin{tabular}{c c c c c} 
  \hline\hline 
  Case           & $\alpha\ind{CGM}$ & $Y\ind i$  & $Z\ind i$   & $\KW$ (cgs) \\ 
  \hline 
  GN93, no b.  & 0.7031   & 0.2792 & 0.02101 & 0 \\ 
  GN93, b.     & 0.6842   & 0.2760 & 0.02019 & $6.2\times 10^{47}$ \\
  AGS05, no b. & 0.6702   & 0.2607 & 0.01371 & 0 \\
  AGS05, b.    & 0.6502   & 0.2566 & 0.01307 & $5.6\times 10^{47}$  \\
\hline 
\end{tabular}
\tablefoot{$\alpha\ind{CGM}$ is the mixing length, $Y\ind i$, $Z\ind i$ the initial helium and metallicity abundances,
  and $\KW$ the braking parameter  entering Eq. (\ref{eq:kawaler}). No b. indicates no braking, b. braking.}
\end{table}

\begin{table}
 \caption{Characteristics of calibrated solar models.
} 
 \label{tab:sunres} 
 \centering 
 \begin{tabular}{c c c c c} 
  \hline\hline 
  Case           & $Y\ind s$  & $Z\ind s$ & $r\ind{CZ}/R_{\odot}$ \\ 
  \hline 
  GN93, no braking  & 0.2511 & 0.01950 & 0.7123 \\ 
  GN93, braking     & 0.2613 & 0.01950 & 0.7166 \\
  AGS05, no braking & 0.2319 & 0.01260 & 0.7314 \\
  AGS05, braking    & 0.2425 & 0.01260 & 0.7364 \\
\hline 
\end{tabular}
\tablefoot{ $Y\ind s$ and $Z\ind s$ are the
surface helium abundance and metallicity, respectively, and $r\ind{CZ} /R_{\odot}$ is the normalized 
radius at the base of the convective envelope.}
\end{table}

\section{Comparison with results from other evolutionary codes}\label{sec:compare}%

The complicated nature of the equations describing the evolution of the rotation profile (and particularly
the evolution of $U_2$) made it important to validate our approach. 
To do that we compared our results  with those  obtained with other implementations of rotational mixing in stellar
 evolution codes, namely STAREVOL \citep{2003A&A...399..603P} 
for the solar case \citep[see also][]{2010ApJ...715.1539T} and the Geneva stellar evolution code 
\citep{1997A&A...322..209T}
 for the case of $3 M_{\odot}$ and $5 \msol$ stars.

\subsection{Standard physics}

In these comparisons, we used the OPAL equation of state \citep{Rogersetal:1996} and opacities \citep{Iglesias/Rogers:1996}, 
complemented at $T<10^4{\rm K}$ by the \citet{Alexander/Ferguson:1994} opacities. We used the NACRE nuclear reaction 
rates of \citet{Anguloetal:1999} except for the $\element[][14]{N}+p$ reaction, for which we used the reaction rates given in
 \citet{2004A&A...420..625I}. The solar compositions of \citet[hereafter GN93]{1993PhST...47..133G} and 
\citet[hereafter AGS05]{2005ASPC..336...25A} were used.
 The Schwarzschild criterion was used 
 to determine convective instability. Convective core overshoot fully mixes the chemical composition to a
 distance $d\ind{ov} = \alpha\ind{ov} \times {\rm min} \left( \Hp, r\ind{co} \right)$ from the border of the convective 
core (where $r\ind{co}$ is the radius of the core determined by  the Schwarzschild criterion). The temperature gradient in the 
overshoot zone is $\nabla=\nabla\ind{ad}$. 
The centrifugal acceleration is taken into account by adding the average centrifugal acceleration $2 \Omega^2 r/3$ 
to gravity in the hydrostatic equilibrium equation.
The atmosphere is computed in the gray approximation and integrated up to an optical depth of $\tau=10^{-4}$.

\begin{figure}[t]
\centering
 \includegraphics[height=8cm]{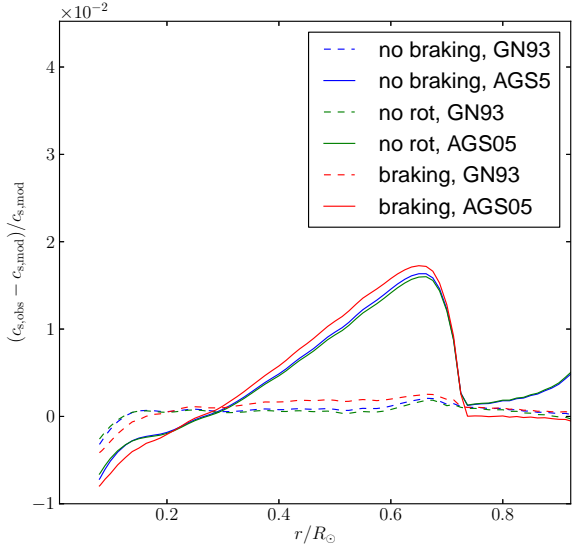}
\caption{Differences in sound speed profiles between  
 the seismic models of \cite{2001ApJ...555L..69T}  and CESTAM.
  }\label{fig:sun}
\end{figure}

\subsection{The Sun} \label{sec:sun}

We computed calibrated rotating solar models using the compositions of GN93 and
AGS05, with and without magnetic braking according to 
the magnetic braking law, Eq. (\ref{eq:kawaler}). 
Models were calibrated to within $10^{-5}$ in luminosity, radius, and surface metallicity 
($Z=0.0195$ for GN93, $Z=0.0126$ for AGS05). 
Initial rotational velocities (and parameter $\KW$ for the models with magnetic braking) 
were chosen so that models have an equatorial velocity $v\ind{eq}=2.02$ km\,s$^{-1}$ at the solar age (4.6 Gyrs). 

The temperature gradient in convective zones was computed using  the convection model of 
\citet{Canutoetal:1996} with $l = \alpha_{\rm CGM} \Hp$. 
All models include microscopic diffusion and settling using the approximations proposed by \citet{Michaud/Proffitt:1993}. 
The initial rotational velocities for the cases with braking were chosen in order to have $v\ind{eq}=20$ km\,s$^{-1}$  at the ZAMS.

Table \ref{tab:sunparams} shows the values of parameters resulting from the solar calibration in $\alpha\ind{CGM}$, the 
initial helium, and metallicity abundances ($Y\ind i$ and $Z\ind i$). The greatest differences
 between the cases with and without magnetic braking concern the initial abundances. This is because models
 with braking rotate much faster at the center \citep[as shown in][]{2010ApJ...715.1539T} and thus have a higher
 $\Omega$-gradient, leading to a much higher turbulent diffusion. Turbulent diffusion tends to homogenize the radiative zone to a large
extent, partially erasing the composition gradient created by gravitational settling.

Table \ref{tab:sunres} shows some characteristics of the calibrated solar models. The higher turbulent diffusion partially 
stops the settling of helium, causing a higher helium abundance in the convective zone. 
However, the strong $\Omega$-gradient predicted by models with braking in the radiative zone is in direct contradiction with helioseismic 
results, which indicate a flat rotation profile to within $r\simeq 0.25 R_{\odot}$ of the center
 \citep[see, e. g.][]{1999MNRAS.308..405C,2003ApJ...597L..77C,2008ApJ...679.1636E}.
  A new physical mechanism is needed to explain the 
discrepancy, such as transport of angular momentum by internal gravity
 waves \citep[see][]{2005A&A...440..981T} 
and/or magnetic stresses \citep{2004A&A...422..225M, 2005A&A...440..653M}

Figure \ref{fig:sun} shows differences between the sound speed 
  $c\ind s$ profiles of the seismic Sun of \citet{2001ApJ...555L..69T} using results from SoHO (GOLF-MDI) and 
our calibrated solar models \citep[see also][for intermediate opacities AGS09]{2011JPhCS.271a2031G}. 
 The seismic Sun of \citet{2001ApJ...555L..69T} agrees with \citet{2009ApJ...699.1403B} 
using results from BiSON and MDI. A good agreement   at the level of 0.3 \% 
is seen when using the  old mixture of GN93  with only small discrepancies below the convection zone
 and in the central region. On the other hand, severe discrepancies occur 
 when more recent mixtures are used, such as AGS05. 
  These results  are  quite   similar to  what is found in the literature
   \citep[see, e. g., ][ for models with no rotation induced transport]{1997MNRAS.292..243B,2008PhR...457..217B,2011ApJ...731L..29T}. 

We compared our results for calibrated solar models including rotation-induced transport of Type I
with the results obtained in \cite{2010ApJ...715.1539T} with the STAREVOL code. We found  good 
agreement between the profiles of the angular velocity, $U_2$, and the two components of turbulent viscosity. 
Several improvements have been implemented in the code CESTAM compared to the 
CESAM2K 
version used in \cite{2010ApJ...715.1539T}, but with no consequence for the solar case.
 We confirm that rotation induced transport of Type I
does not help remove the discrepancies, as found by \cite{2010ApJ...715.1539T}.

\subsection{Higher mass,  main sequence stellar models}

\begin{figure}[t]
\centering
\includegraphics[height=6cm]{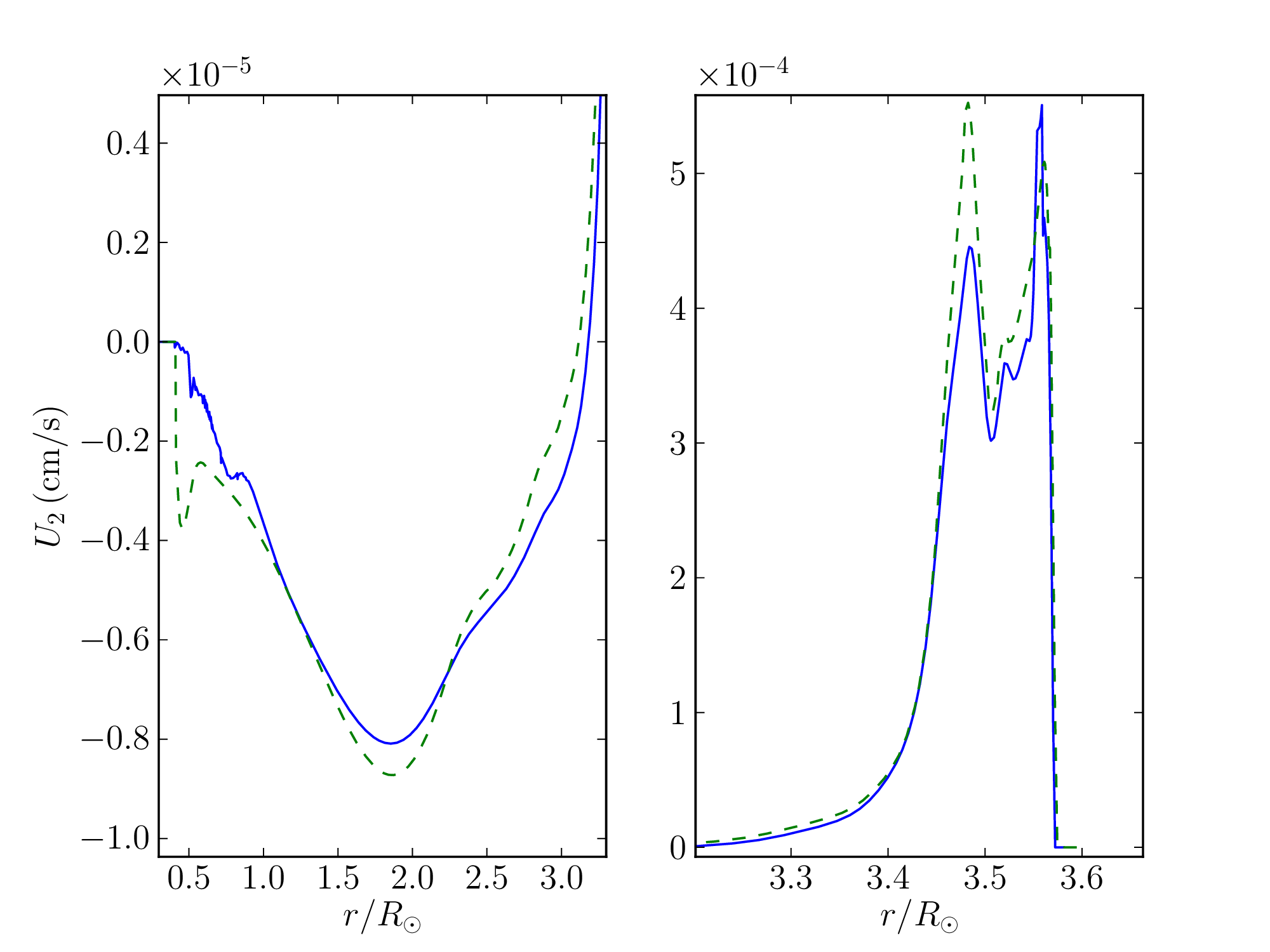}
\caption{Profiles of the vertical component of the meridional circulation $U_2(r)$ as a function of normalized radius ($r/R_\odot$) 
 for  5 $M_\odot$ stellar models computed with CESTAM (continuous line) 
and the Geneva code (dashed line) 
when the central hydrogen abundance is  $X\ind c=0.35$ (see text). 
The left panel shows the central regions, the right panel the surface.}
\label{fig:5msun_u2}
\end{figure}

We compared results  for models 
with  higher mass stars. These stars have thin convective or fully radiative surface layers and are
therefore not expected to undergo  magnetic braking. Calculations here are carried out assuming
global conservation of  angular momentum, which is obtained at the precision level of $10^{-6}$.

The first comparison concerns the  evolution of a $5 M_{\odot}$ model computed with CESTAM and 
Geneva codes \citep[as in][]{1997A&A...322..209T},
 using $X\ind i=0.73$ and $Z\ind i=0.01$ (and no magnetic braking). These models were computed without microscopic diffusion and settling.
 Figure \ref{fig:5msun_u2} shows the profile of $U_2(r)$ at the middle 
 of the MS (when $X\ind c=0.35$).
Curves for the same quantities superimpose, showing excellent agreement between the results obtained with the two codes. There are small differences 
that we attribute to the different microphysics used in the codes.  The rotation profiles, not shown, also agree during the course of the evolution.
\begin{figure}[t]
\centering
\vskip0.5cm
 \includegraphics[height=6cm]{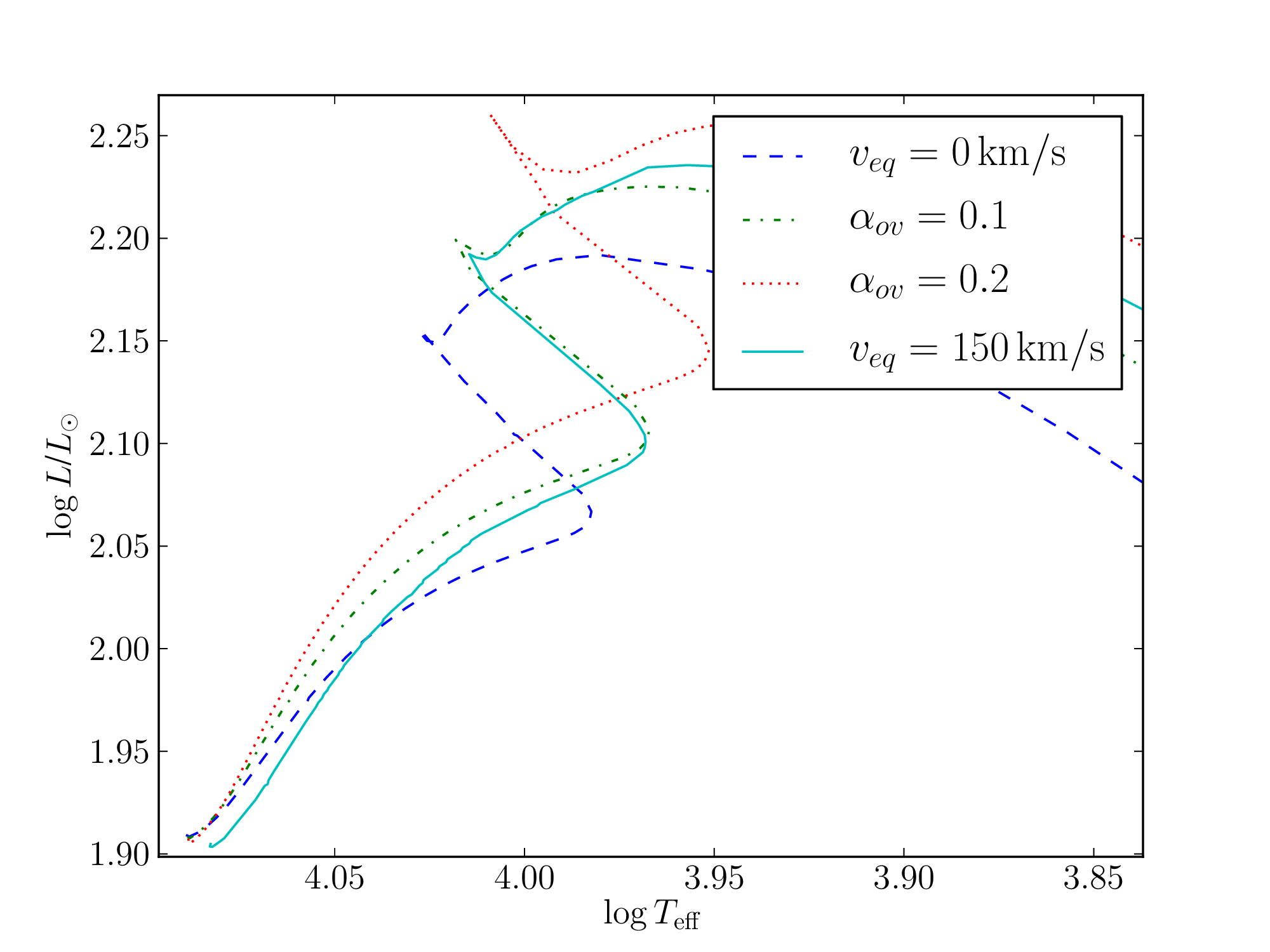}
\caption{Evolutionary tracks on the HRD for a $3M_{\odot}$ model 
calculated with CESTAM code assuming either no extra mixing (dashed) or 
rotation with an initial equatorial velocity $v\ind{eq}=150$\,km\,s$^{-1}$ at the ZAMS (continuous line) or
 without rotation but with overshooting 
  and $\alpha\ind{ov}=0.1$ (dot-dashed line) and $\alpha\ind{ov}=0.2$ (dotted line).}\label{fig:3msun}
\end{figure}

\medskip 

We also compared the effects of rotational mixing on the evolutionary tracks 
on the Herzsprung-Russel diagram (HRD) between CESTAM models
 and those of \citet{2010A&A...509A..72E} computed with  the Geneva code \citep{2008Ap&SS.316...43E}. 
 Figure \ref{fig:3msun} shows evolutionary tracks on the HRD
  for a $3 M_{\odot}$ model calculated without extra mixing (microscopic diffusion, convective overshoot, rotation), 
  with rotational mixing only and with overshooting only.

We considered two cases for the models with overshooting,
 $\alpha\ind{ov}=0.1$ and $\alpha\ind{ov}=0.2$. 
 The model with rotational mixing has $v\ind{eq}=150$\,km\,s$^{-1}$ at the ZAMS. 
 As in \citet{2010A&A...509A..72E}, the evolution of the rotating model 
 closely resembles the evolution with $\alpha\ind{ov}=0.1$. The difference between the rotating models and the nonrotating models 
at the ZAMS is due to the centrifugal acceleration. The decrease in the effective gravity (gravity minus centrifugal force) 
replicates a nonrotating star with a lower mass.

 The evolutionary tracks shown in Fig. \ref{fig:3msun} are identical to those in Fig.~2 of \citet{2010A&A...509A..72E}.

\subsection{Evolved low-mass models} \label{sec:redgiant}

The rotation profiles that we obtain for our red-giant models are similar to those 
computed by \citet{2010A&A...509A..72E}. Examples for $M_{\star}= 1.3 M_\odot$  are given in the section below.

\section{Evolution of rotational splittings from PMS  to RGB}\label{sec:evsplit}%
\begin{figure}
\centerline{
\includegraphics[width=8cm]{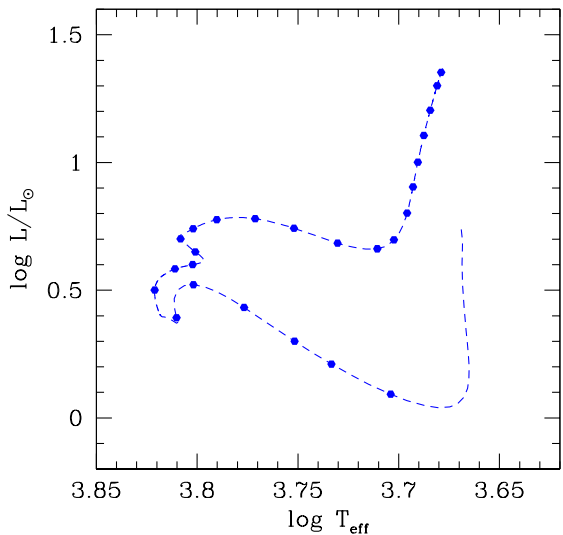}}
\caption{ Evolutionary track for a $1.3 M_\odot$ sequence of models from PMS to
 RGB. Dots indicate models for which rotational splittings are computed. The 
 numbering of the selected models  stars with the first dot on the PMS and ends with number 25 
 as the highest dot on the red-giant branch
  }
\label{fig:HR}
\end{figure}

Rotational splittings are useful as seismic diagnostics   for measuring the
 stellar internal rotational profile. For slow  rotation, a first order perturbation description 
provides rotational splittings that are linearly dependent on  the rotation profile. They are
therefore convenient tools, easy to compute from stellar models and theoretical oscillation codes
for comparison with  observations.  The first order  approximation provides 
the following expression for the linear rotational splittings 
\citep[][and references therein]{1991sia..book..401C}:


\begin{equation}
\delta \nu_{n\ell} =  \int_0^1 K_{n\ell}(x) ~\frac{\Omega(x)}{2\pi}~ \d x 
\label{Eq:split}
\end{equation}
where $x=r/R_{\star}$ is the normalized radius,  $\Omega$ is the angular rotation (rad/s) 
and the rotational kernel $K_{n\ell}$ takes the form
\begin{equation}
 K_{n\ell} =\frac{1}{I_{n\ell}}~
  \Bigl[\xi_r^2+\ell(\ell+1) \xi_h^2-2  \xi_r \xi_h - \xi_h^2 \Bigr] ~\rho~x^2~ ,
\end{equation}
where
 $I_{n\ell}$ is the mode inertia 
\begin{equation}
I_{n\ell}= {\int_0^1 ~\Bigl[\xi_r^2+\ell(\ell+1) \xi_h^2\Bigr]~\rho~x^2~ \d x}.
\end{equation}
The quantities entering the equations above are the fluid vertical and
 horizontal displacement eigenfunctions,
 $\xi_r$ and $\xi_h$ respectively, and the density $\rho$. For solid-body
 rotation, the rotational splittings become:
\begin{equation}
\delta \nu_{n\ell} =  \beta_{n\ell} ~\frac{\Omega}{2\pi}
\label{deltanu}
\end{equation}
with
\begin{equation}
 \beta_{n\ell} = 1-C_{n\ell}= \frac{1}{I_{n\ell}} \int_0^1 \Bigl[\xi_r^2+\ell(\ell+1) \xi_h^2-2  \xi_r \xi_h - \xi_h^2 \Bigr] 
 ~\rho~x^2~ \d x,
\end{equation}
where $C_{n\ell}$ are the Ledoux coefficients.

For asymptotic  pure p-modes, $C_{n\ell} \sim 0$ and $\beta_{n\ell}\sim 1$. For pure g-modes,  
 $C_{n\ell} \sim 1/\ell(\ell+1)$ and $\beta_{n\ell}\sim  1-1/\ell(\ell+1)$, i. e. $\beta_{n\ell}\sim  1/2$ for $\ell=1$ modes.

\begin{figure}
\centerline{
\includegraphics[width=9cm]{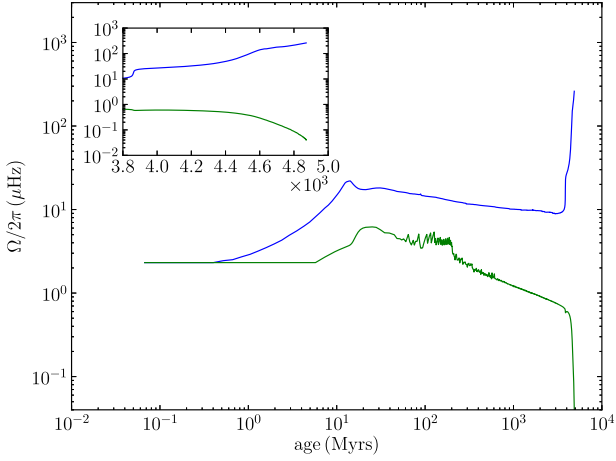}}
\caption{ Evolution of the central (blue line) and surface (green line) rotation rates  for a $1.3 M_\odot$
 sequence of models from the PMS to the
 RGB.  The insert shows the central and surface rotation rates along
   the red-giant branch. }
\label{fig:rottot}
\end{figure}

 To  investigate 
  the evolution of rotational splittings  with 
  stellar age, we computed  a 1.3 $M_\odot$ evolutionary sequence
  including rotationally induced transport as described in Section~\ref{sec:physinps} above.
   We used the same input physics as in the case AGS05 with  
braking described in Section 
\ref{sec:sun}, but without microscopic diffusion and settling, as we found that microscopic diffusion and settling
 do not affect the rotation profiles of our models.  We chose $M_{\star} = 1.3 M_{\odot}$ because it is 
a typical seismic mass obtained for the \Kepler and CoRoT red-giant stars.

Figure~\ref{fig:HR}  shows an evolutionary track for a sequence of  $1.3 M_\odot$ stellar models 
 in a HR diagram computed with CESTAM. 
In order to interpret the rotational splittings, we plot the evolution of the  central and
surface rotation rates with age in Fig.~\ref{fig:rottot}.  

 We then computed oscillation frequencies and rotational splittings for several models spanning 
the track (shown in Fig.~\ref{fig:HR})  using  the freely available ADIPLS adiabatic oscillation code \citep{2008Ap&SS.316..113C}. 
We computed frequencies of $\ell=0,1$  axisymmetric modes and rotational splittings 
for $\ell=1$ modes.  
 
The $1.3 M_\odot$ models have a
convective envelope from PMS to the RGB, hence we expect   stochastically excited modes all along the
sequence.  Thus, we chose the frequency range that spans an
interval of a few  radial orders below and above  $n\ind{max}$, the radial order corresponding to the
frequency at maximum power.   It is estimated as 
$n\ind{max} = \nu\ind{max}/\Delta \nu$, where the frequency at maximum power spectrum $\nu\ind{max}$ 
and the mean large separation $\Delta \nu$ are given by the usual scalings relations 
\begin{eqnarray}
\nu\ind{max}  &= & \nu_{{\rm max},\odot}  ~\Bigl(\frac{M_{\star}}{M_\odot} \Bigr)~\Bigl(\frac{R_{\star}}{R_\odot}\Bigr)^{-2}\Bigl(\frac{T\ind{eff}}{T_{{\rm eff},\odot}} \Bigr)^{-1/2}\\
\Delta \nu &=&  \Delta \nu_{\odot} ~ \Bigl(\frac{M_{\star}}{M_\odot}\Bigr)^{1/2}~\Bigl(\frac{R_{\star}}{R_\odot} \Bigr)^{-3/2}
\end{eqnarray}
with $T_{{\rm eff},\odot}=5777 K$, $\nu\ind{max} = 3050\mu$Hz  and $\Delta \nu_{\odot}=134.7 \mu$Hz for the Sun.
Indeed, it has been conjectured, then shown observationally, that these relations 
predict well the location of the excited frequency range of
stochastically excited modes \citep[e g. ][]{1991ApJ...368..599B, 1995A&A...293...87K, 2010A&A...509A..77K}


\subsection{Validity of  linear approximation for the rotation splittings}

\begin{figure}
\centering
\includegraphics[width=7cm]{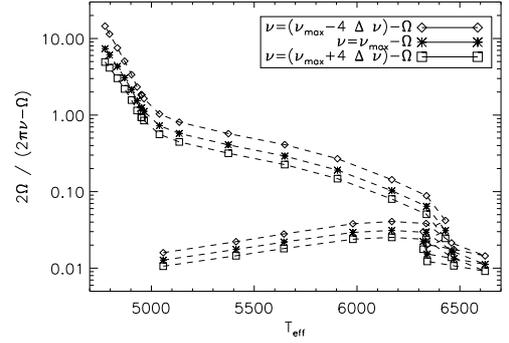}
\includegraphics[width=7cm]{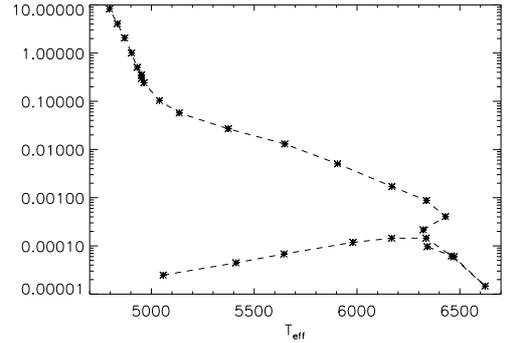}
\caption{The  evolution of the parameters $\zeta$, Eq.~(\ref{zeta}) (top)  and $\Delta$, 
Eq.~(\ref{delta}) (bottom), with the effective temperature of the
model along its evolution from the PMS to the red-giant branch. To be conservative, the layer 
where $\Omega$ is maximum is considered for $\zeta$ and the layer where the gradient is maximum 
is chosen for computing $\Delta$.  For $\zeta$, we used $m=-1$, since $\zeta$ is higher for prograde modes,
 while for $\Delta$ we considered $m=1$ as no difference 
is found for $m=-1$ due to the fact that $\Omega << \omega$ where $\Delta$ is maximum. }
\label{validity}
\end{figure}

The validity of the linear approximation is estimated by comparing the rotation rate to the 
oscillation frequency. In a perturbative approach, the parameter 
\begin{equation}
\zeta \equiv  \frac{2 \Omega}{\omega\pm \Omega}
\label{zeta}
\end{equation}
 is thus assumed  smaller than
unity. We evaluate this parameter at the center of each selected equilibrium 
 model of the evolutionary sequence,  where $\Omega$ is largest, and for a
 range of frequency spanning the radial order interval $(n\ind{max}-4,n\ind{max}+4)$. 

 Sharp rotation
 gradients develop  in the central layers and can also cause some departure from the
 linear approximation. In the oscillation equations \citep[see, e. g. Eqs. (34.7-34.12) in][]{1989nos..book.....U},
the rotation gradient term appears with a factor 
\begin{equation}
  \frac{1}{\left( \omega \pm \Omega\right)^2} \frac{\d \Omega^2}{\d \ln r}.
\end{equation}

 Thus, we consider that the linear approximation is
valid if
\begin{equation}
 \frac{1}{\left( \omega \pm \Omega\right)^2} \frac{\d \Omega^2}{\d \ln r} = \zeta \frac{1}{\omega \pm \Omega}\frac{\d \Omega}{\d \ln r} <
 \frac{1}{\omega \pm \Omega}\frac{\d \Omega}{\d \ln r} << 1,
\end{equation}
where the first inequality follows from assuming $\zeta < 1$. 

We then evaluate the quantity
\begin{equation} 
 \Delta \equiv \Bigl( \frac{\d\Omega}{\d\ln r}\Bigr) \frac{1}{\omega\pm\Omega}
 \label{delta}
\end{equation}   at the radius where 
 the rotation gradient is greatest   and for  $\omega/2\pi =\nu\ind{max}$.
 Fig.~\ref{validity}  shows both quantities, $\zeta$ and $\Delta$, 
 as functions of the effective temperature of the models. Both parameters remain much smaller than
 unity until  the model reaches the base
 of the red-giant branch. Therefore, the use of the linear approximation for the rotational splittings can
 be safely used for stochastically excited solar-type modes of low mass stars 
  except  for the fast rotating cores of red giants.  
 
\subsection{Evolution of rotational splittings  of stochastically excited modes along an
evolutionary track}

\begin{figure}[ht]
\centering
\includegraphics[width=7cm]{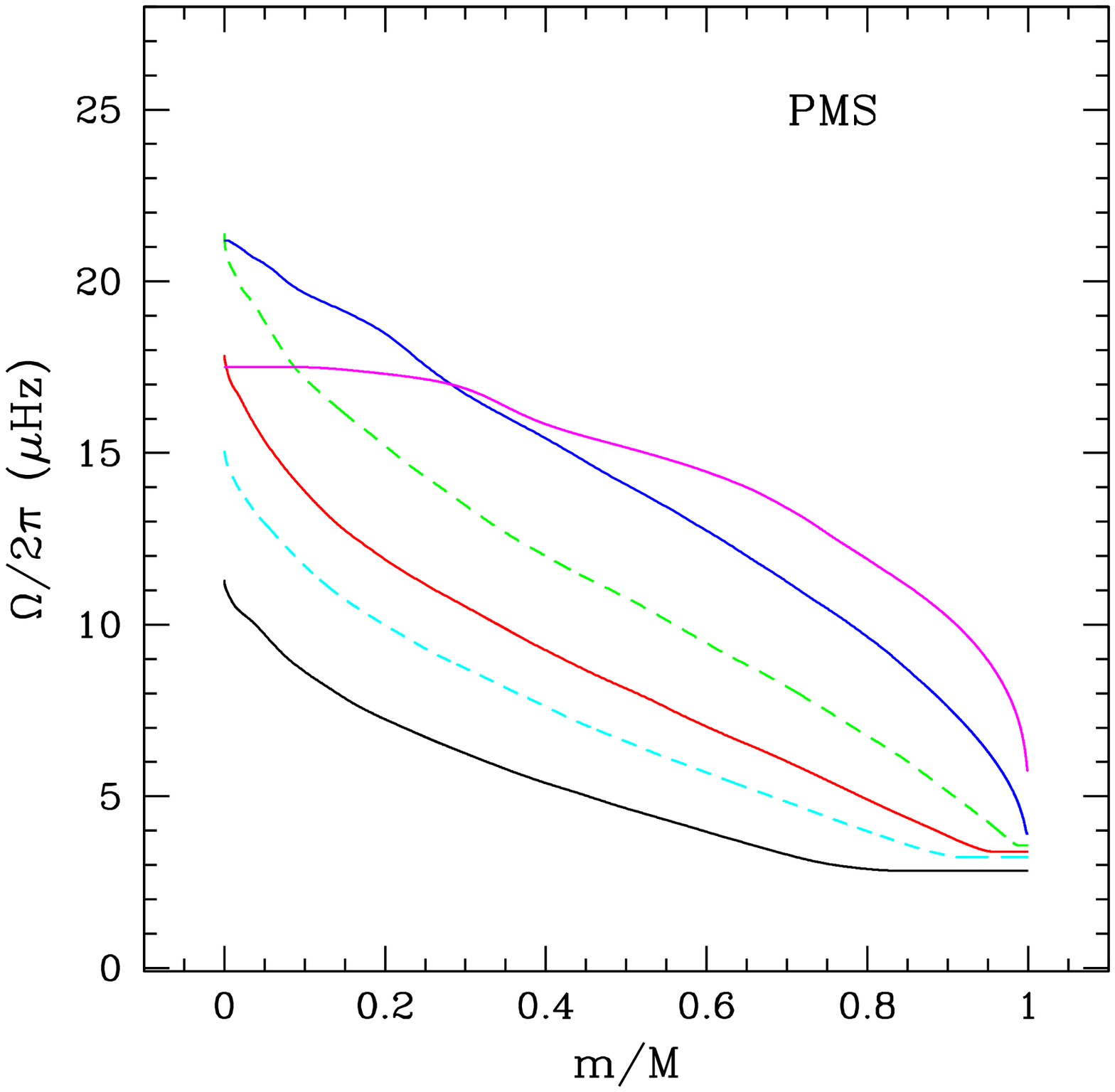}
\includegraphics[width=7cm]{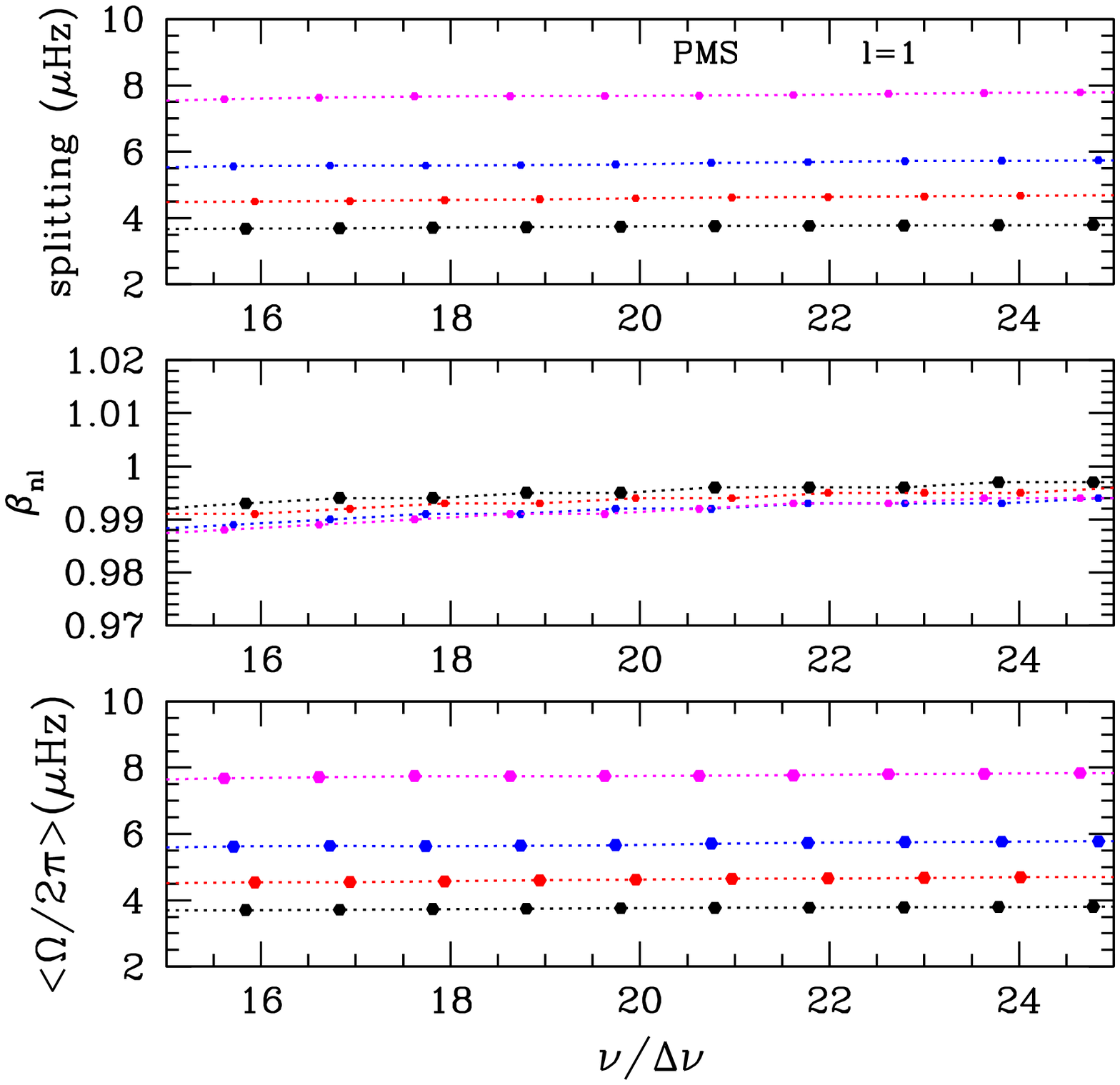}
\caption{  {\it Top panel:} Rotation profiles as a function of the normalized mass $m/M_{\star}$ 
 for $1.3 M_\odot$ PMS models \# 1 (black), \# 2 (cyan), \# 3 (red), \# 4 (green), \# 5 (blue) and \#6 (magenta)  shown in Fig.~\ref{fig:HR}.
 {\it Second panel:} Rotational splittings for $\ell=1$ modes as a function of the normalized frequency 
 $\nu/\Delta \nu$ for models \# 1,\# 3,\# 5 and \#6 (same colors as above).
  {\it  Third panel:} $\beta_{n\ell}$  for the same models. 
  {\it  Bottom panel:} $\langle \Omega/2\pi \rangle =\delta \nu_{n\ell}/\beta_{n\ell}$ for the same models. 
The large separation $\Delta \nu$ goes from $87.9\,\mu$Hz (model \# 1) to $106.2\,\mu$Hz (model \# 6).}
\label{rotPMS}
\end{figure}

 In the following, we discuss the information that can be retrieved from the average value throughout the star 
 $\langle \Omega/2\pi \rangle = \delta \nu_{n\ell} /\beta_{n\ell}$ weighted by the mode.
This definition is given at fixed $\ell$, and
depends on the radial order $n$. 
\medskip

\begin{itemize}
\item {\it The PMS regime:} 
On the PMS,  the central and surface rotation rates are equal as long as the 
model is fully convective.  The rotation rate remains constant in time  for the duration of  disk locking. 
When the radiative core appears, the central rotation starts to increase due  to the
contraction of the central layers. When disk locking stops, the
 surface rotation rate starts increasing as the star contracts.
Fig.~\ref{rotPMS} shows the rotation profiles of the selected PMS  models shown
 in Fig.~\ref{fig:HR}. 
The surface rotation  rate evolves from $2.84 ~\mu {\rm Hz}$ (model \# 1)  to $3.9 ~\mu {\rm Hz}$ 
(model \# 5) while the  central rotation rate increases from $11.3 ~\mu {\rm Hz}$  to $21.2 ~\mu {\rm Hz}$.
 The  rotational splittings for $\ell=1$ modes are computed  for models \# 1 to \# 6 
 according to Eq.~(\ref{Eq:split}) and 
 are also shown in  Fig.~\ref{rotPMS}. The splittings are nearly 
 independent on radial order and increase steadily with the surface rotation rate
as the model evolves along the PMS towards the ZAMS.
 The value of the Ledoux constant almost vanishes, and $\beta_{n\ell}\sim 1$, 
 because the modes are  essentially pressure modes  in the frequency range where they are expected to be
stochastically excited. These modes have no amplitudes in the
inner layers and cannot probe the rotation there. Thus the mean rotation rate $\langle \Omega \rangle/2\pi$
 corresponds essentially
to the rotation averaged over the outer
layers.  As the rotation increases inwards in these layers, 
we find that $\langle \Omega \rangle/ 2\pi = 3.8~\mu {\rm Hz}$  for Model \# 1, close to -  but 
slightly larger than - its surface rotation rate.  
Model \# 6 is already on the MS. 
 Its surface rotation is faster than that of
 younger models but its central rotation is slower. 
This is not seen in the  corresponding splittings
and the averaged rotation $\langle \Omega \rangle/ 2\pi$ remains close to the surface value.
 Results are similar for $\ell=2$ modes (not shown).  

   \begin{figure}[t]
   \centering
\includegraphics[width=7cm]{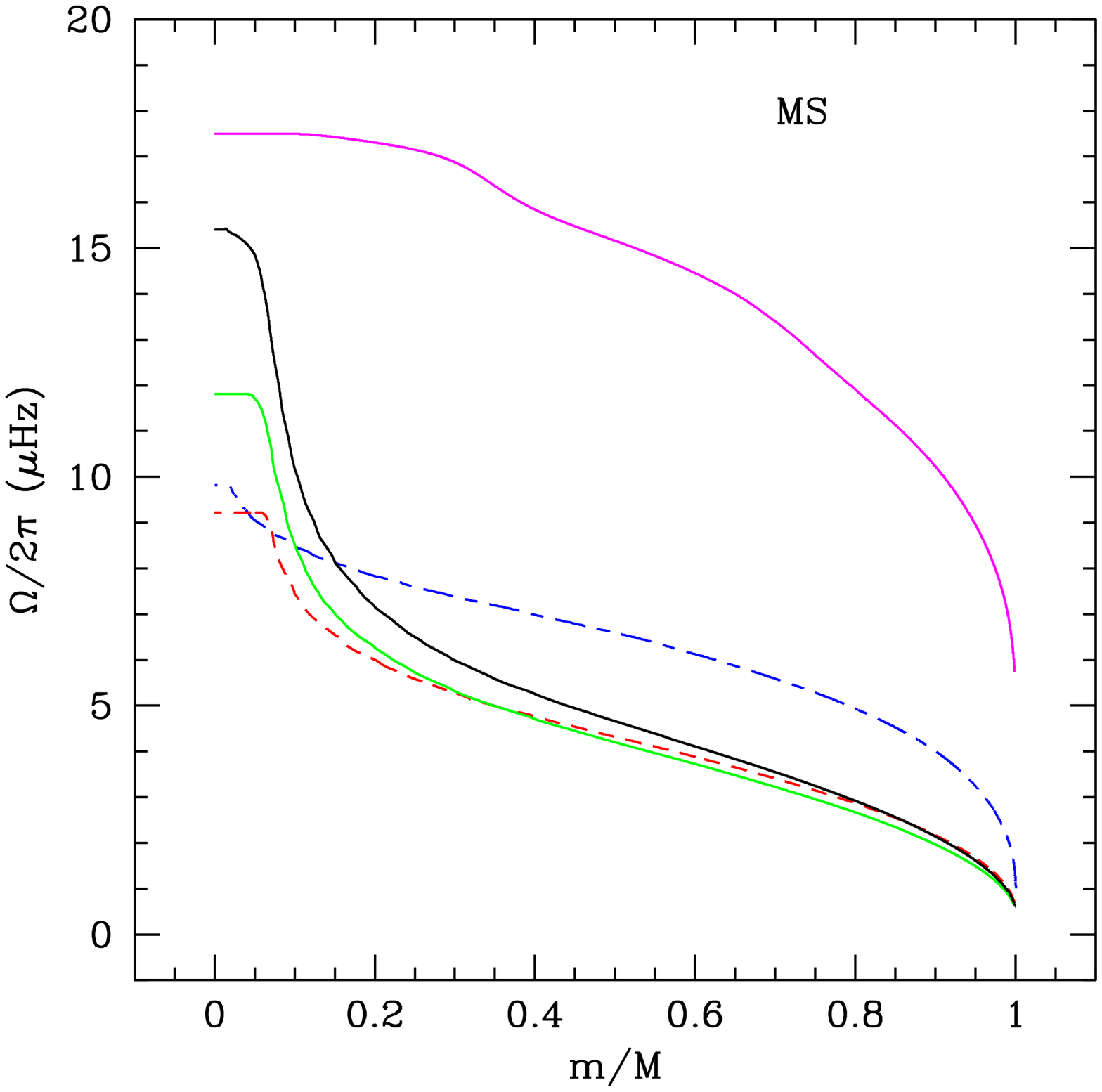}
   \includegraphics[width=7cm]{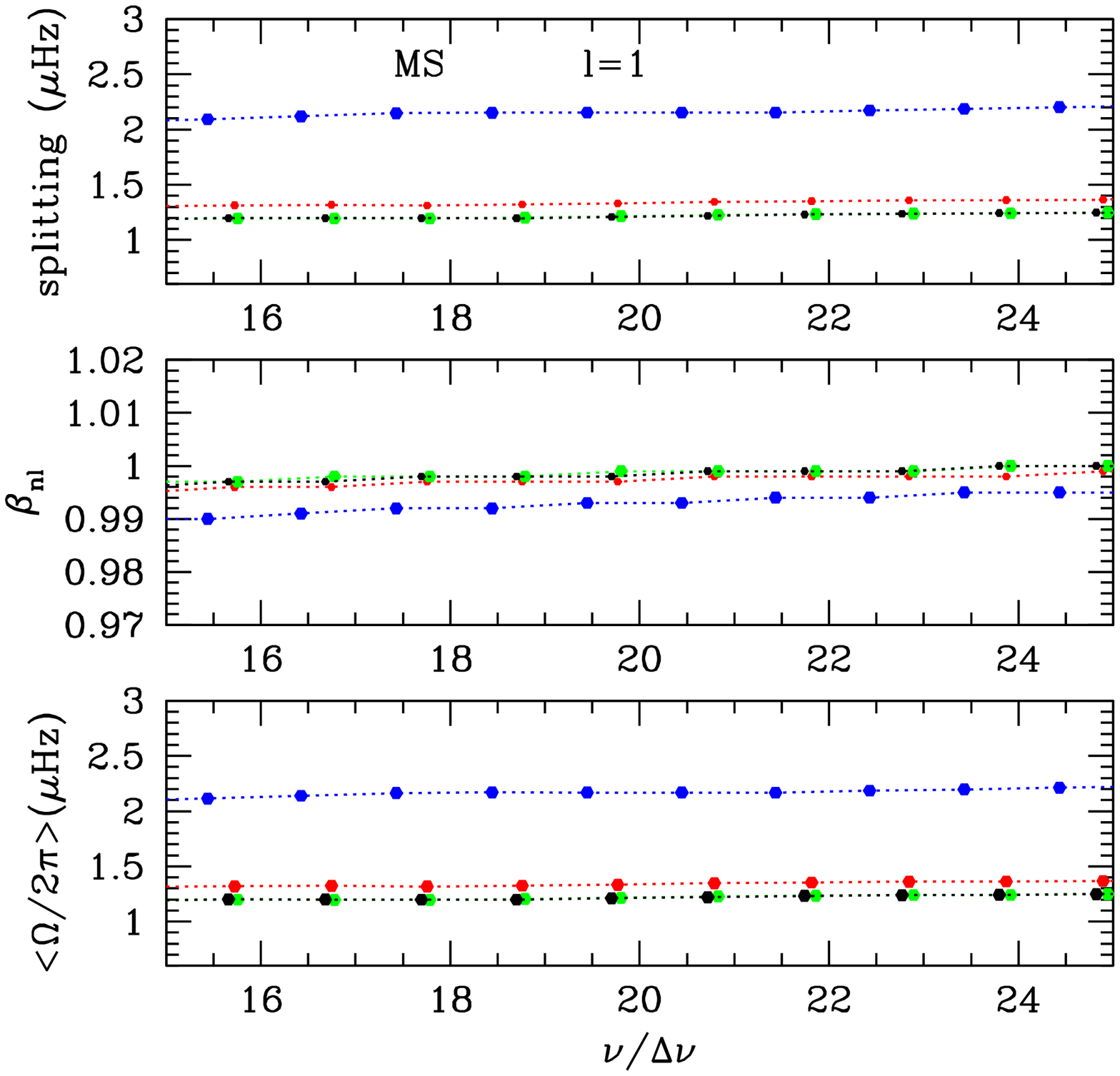}
 \caption{ {\it Top panel:} Rotation profiles as a function of the normalized mass $m/M_{\star}$ 
 for $1.3 M_\odot$ MS models \# 6 (magenta), \# 7 (blue), \# 9 (red), \#10 (green) and \# 11 (black)  
  shown in Fig. \ref{fig:HR}.
 {\it Second panel:} Rotational splittings for $\ell=1$ modes as a function of the normalized frequency 
 $\nu/\Delta \nu$ for models \# 7,\# 9,\# 10 and \# 11 (same colors as above).
  {\it  Third panel:} $\beta_{n\ell}$  for the same models. 
  {\it  Bottom panel:} $\langle \Omega \rangle=\delta \nu_{n\ell}/\beta_{n\ell}$ for the same models.
The large separation $\Delta \nu$ goes from $97.6\,\mu$Hz (model \# 7) to $63.2\,\mu$Hz (model \# 11).}
    \label{fig:rotMS}
   \end{figure}

\medskip

\item {\it The MS regime:} 
On the MS,  the surface rotation rate decreases with age due to  both braking at the surface and an increase of the stellar
radius. The central rotation rate also decreases with time due to transport of angular momentum
from the core to the surface.
The rotation  in the central regions
 first decreases from Model \#6 to \#9 and  increases by roughly 60 \% 
 from model  \#9 to  \#11 (Fig.~\ref{fig:rotMS}).  
The excited modes are still in the frequency domain of high order p-modes and their rotational splittings 
 reflect the surface behavior only. The splittings decrease with the surface angular velocity.
At the end of the MS (model \#11), 
 the central rotation rate has already reached 15.4 $\mu$Hz,
 while the surface rotates at a rate of 0.6 $\mu$Hz: the core is rotating roughly 
 26 times faster than the surface. As can be seen in Fig. \ref{fig:rotMS}, the rotational
 splittings for this model yield a mean rotation  $\langle \Omega \rangle/ 2\pi= 3.8~\mu {\rm Hz}$, that
 is 1.22 $\mu{\rm Hz}$ higher than the surface rotation rate
 but much lower than the central rotation rate.

\begin{figure}
\centering
\includegraphics[width=7cm]{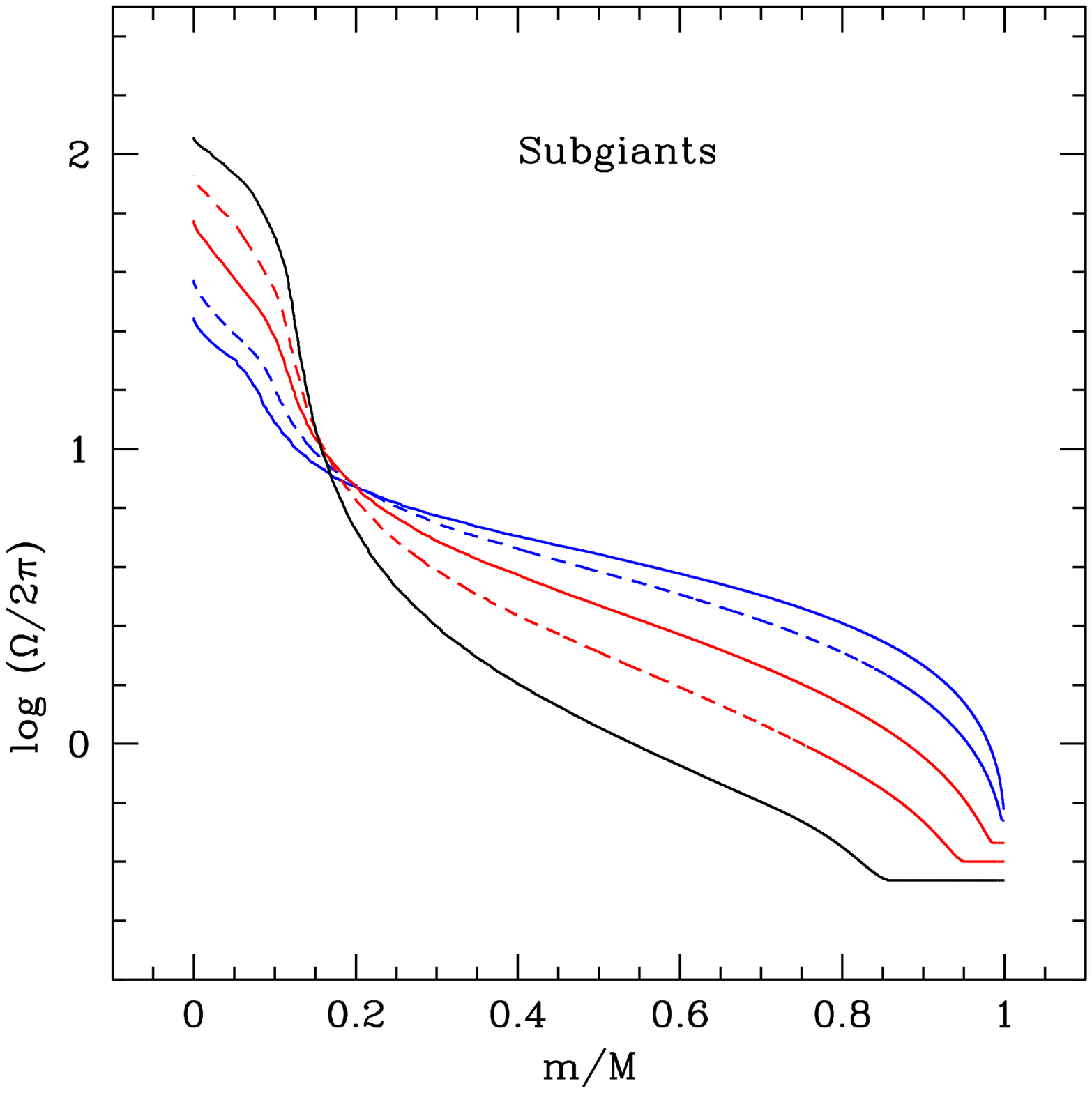}
\includegraphics[width=7cm]{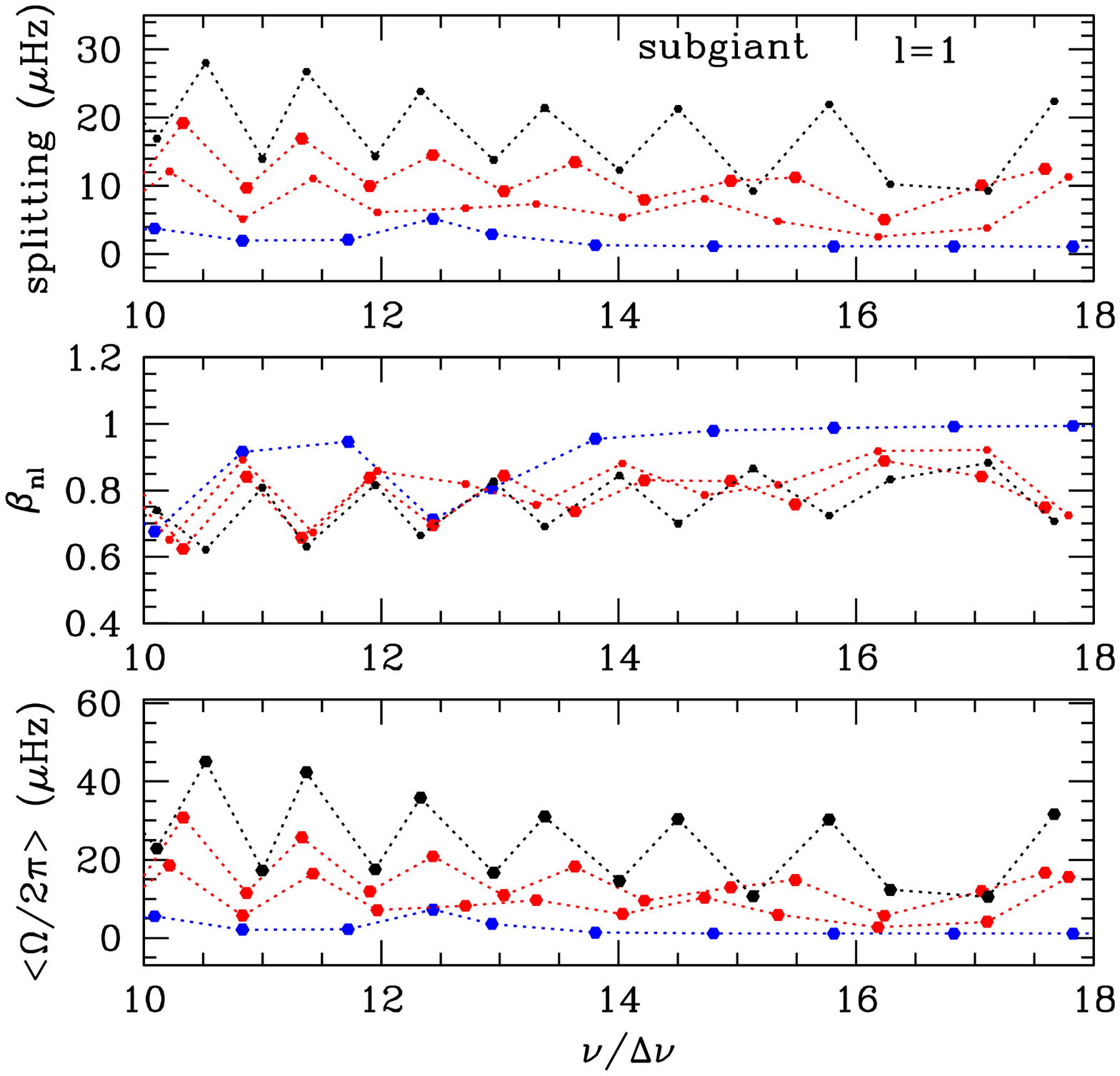}
\caption{ {\it Top panel:} Rotation profiles (logarithmic scale) as a function of the normalized mass $m/M$ 
 for $1.3 M_\odot$ subgiant  models \# 12  (blue solid) to \# 17 (black solid) 
   as indicated in Fig. \ref{fig:HR}.   
 {\it Second panel:} Rotational splittings for $l=1$ modes as a function of the normalized frequency 
 $\nu/\Delta \nu$ for model \#12,\#14,\#15 and \#16.
  {\it  Third panel:} $\beta_{n\ell}$  for the same models. 
  {\it  Bottom panel:} $\langle \Omega \rangle/2\pi=\delta \nu_{n\ell}/\beta_{n\ell}$ for the same models.
The large separation $\Delta \nu$ goes from $56.5\,\mu$Hz (model \# 12) to $38.0\,\mu$Hz (model \# 16).
 }
\label{fig:rotsubgiant}
\end{figure}

\medskip

\item {\it The subgiant  regime:} 
When the model cools along the subgiant branch, its rotation  keeps evolving 
differently in the inner and outer parts, slowing down at the surface and accelerating in the
 inner regions (Fig. \ref{fig:rotsubgiant}).  Substantial changes appear when the model leaves the 
MS and  evolves as a subgiant. The Brunt-V\"ais\"al\"a 
frequency of red-giant stars is very high in the radiative interior, 
 and the frequencies of g-modes enter the 
frequency domain of p-modes in the frequency range
 where stochastically excited modes are expected \citep{2011arXiv1110.5012C}. 
  During the subgiant phase, avoided crossings between p- and g-modes appear. 
As observationally shown \citep{2010A&A...515A..87D,2012A&A...540A.143M}, 
some  p-modes become mixed modes.
  These modes have amplitudes both in the inner regions,
   where they behave  as gravity modes, and in the surface layers, where they behave as acoustic modes 
  \citep{1991A&A...248L..11D}.
They are very interesting as they allow probing regions deep
 inside the star.
 The number of such mixed modes is small at the beginning of the subgiant branch and increases as
 the star evolves toward the  red-giant branch. 
During this phase, modes will eventually undergo avoided crossings, exchanging nature from 
p to g.


\begin{figure}
\includegraphics[width=8cm]{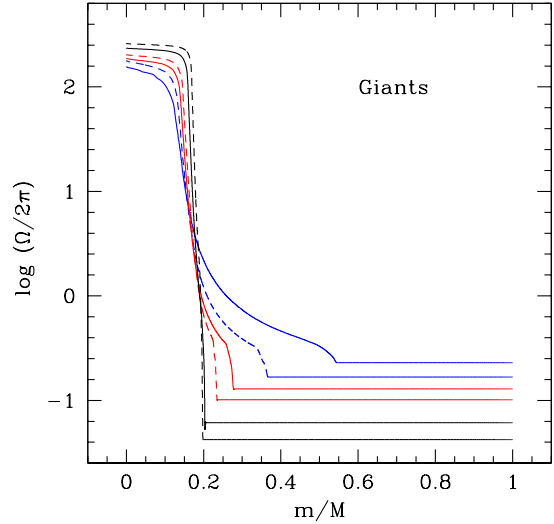}
\caption{Rotation profiles as a function of the normalized mass $m/M_{\star}$
 for  $1.3 M_\odot$  models  \# 18 to \# 21 and \# 23, \# 25
  (from blue to black)  on the red giant
  branch as indicated in Fig. \ref{fig:HR}. }
 \label{fig:rotgiant}
\end{figure}

Figure \ref{fig:rotsubgiant} shows that Model \#12 is not evolved enough so that no mixed modes are 
present in the frequency domain shown. The
splittings remain nearly frequency independent and provide a mean rotation rate close to its surface
value. The variation in the splittings with frequency for the more evolved models \# 13 to \# 17  clearly shows the
presence of mixed modes.  
With amplitudes in the dense interior, mixed modes have a much stronger inertia 
  than the neighboring p-modes \citep{2009A&A...506...57D}. 
 For the same reason, the  rotational  splittings  exhibit the same behavior as the mode inertia although in a
 more pronounced way.  They 
 vary from one mode to the next. They are also larger than for pure p-modes when the rotation increases inward within the model. 
 For subgiant stars, only one  $\ell=1$ mixed mode exists between two consecutive radial modes. 
This explains the saw-like aspect of the variation in the splittings with frequency.
This is confirmed by the same variation in $\beta_{n\ell}$ with frequency: p-modes  have
$\beta_{n\ell}$  close to 1 while mixed modes have $\beta_{n\ell}$  close to 1/2. 
Mixed modes enter the frequency domain by the
low part and one can see that $\beta_{n\ell}$ is closer to 1 at high frequency 
where the g nature of the mixed modes is less pronounced. 
Because of the weighting by the eigenfunction, the values of the splittings yield
a mean rotation $\langle \Omega \rangle/2\pi$ that  is  no longer dominated by the outer layer contribution, 
since the mixed modes have amplitudes in the central region where the rotation rate is larger. 
As a result, one obtains a rotation rate that is  higher than the surface value 
but still much lower than the rotation rate
 of the central regions. 

\medskip 

\item {\it The red-giant branch:}
When the outer convective region progresses inward, its  uniform rotation extends deeper 
as well. The surface rotation rate  decreases with time essentially due to an increase in the stellar  radius.
The core accelerates, and in the intermediate region,  where the H-burning shell lies, a sharp rotation
gradient develops \citep[see][]{2006A&A...453..261P}.  
The frequency spectrum of red giants is composed of g-dominated 
 mixed modes and  mixed modes  with  nearly equal p and g character.
  When the star evolves  up the red-giant branch, the number of g-dominated modes increases 
and largely outnumbers the p-g mixed modes \citep{2009A&A...506...57D}.  
 These modes mostly probe
 the core rotation \citep{2012Natur.481...55B,2012ApJ...756...19D, 2012arXiv1209.3336M}.
 For  model \# 25 ( the last shown in Fig.~\ref{fig:HR}),  the core rotation amounts to $251 \mu$Hz and the expected 
 excited frequencies are  about  $90~\mu$Hz. 
 This indicates a ratio $\zeta=5.6 $ (Eq.~\ref{zeta}).  
 
 At such rotation rates, the  first-order
 perturbation   is not relevant to computing the rotational splittings. Nonperturbative methods must
 then be used. 

However, observations show that several RGB stars have core rotation rates
that are much lower than predicted by our models.
\citep{2012arXiv1209.3336M}. Indeed the observed values amount to a few hundred nHz, whereas here we
find several dozen  to a few hundred $\mu$Hz. This  leads to the  conclusion that the central regions of our
models  rotate too fast and  that rotationally induced transport of the kind
 included in our models is not efficient enough 
 to slow down the central  rotation of red-giant stars. 
 \end{itemize}

\section{Seismic tests of transport of angular momentum: slowing down  the rotating core of red-giant stars}\label{sec:discussion}%

As in the solar case, one can wonder whether another mechanism, such as internal gravity waves or magnetic fields, operates
more efficiently to shape the rotation profile, either 
at the red-giant phase or before.  However, several issues must be discussed before 
we can reach such a conclusion.

\subsection{Uncertainties on stellar modeling and transport of angular momentum}

A possible reason  for  this discrepancy is 
some inaccuracy in the current physical description of stellar models. 
Indeed  several assumptions enter the rotationally induced transport as 
prescribed by \citet{1992A&A...265..115Z} and \citet{1998A&A...334.1000M}. Besides, several other 
uncertainties on the input physics of stellar models can also affect the resulting rotation 
profile
 to some extent. 
In what follows, we discuss the sensitivity of the rotation profile of red-giant models to several uncertainties.

\subsubsection{ Effect of convective overshoot}

A $1.3M_{\odot}$ star has a convective core during the MS.  
Overshoot shifts the tracks on the HR diagram, 
imitating tracks with higher masses. 
As a result, at a given radius in the RGB, the core of the model with overshoot has had less time to spin up, since
it started its contraction later in the evolution. 
An overshoot of $0.1 H_P$ reduces the core rotation rate by about 32\,\% when $R_{\star} = 3.73 R_{\odot}$.

On the other hand, we need to reduce the mass of the model with overshoot by $0.04M_{\odot}$ to reproduce the
RGB at the same location in the HR diagram. In that case, the model with
 overshoot and $M_{\star} = 1.26M_{\odot}$ has a central rotation rate (again, when $R_{\star} = 3.73 R_{\odot}$) that is
 similar to the central rotation rate of the model
without overshoot and $M_{\star} = 1.3M_{\odot}$.

\subsubsection{ Effect of the initial rotation state}

It is well known that, after the early stages on the MS, the surface rotation rate does not depend on the 
initial conditions; it only depends on the magnetic braking law (Eqs. \ref{eq:kawaler} and \ref{eq:reiners};  see, e. g.
\citealt{2009pfer.book.....M} and references therein).
 We found that it is indeed the case for our models. 
The internal rotation profile down to the center is also independent of the initial conditions after the early stages of the MS.

\subsubsection{Impact of the magnetic braking law }

We computed an evolution with the braking law of K88 with a value of $\KW$ twice the solar calibrated 
value to reduce the rotation rate. We found that the surface rotation
rate is slowed down, as expected, but the central rotation rate changes only slightly. For instance, at the TAMS, the surface rotation rate changes 
by 20\,\%, and by 3\,\% at the base of the RGB, while the central rotation rate changes by 2\,\% at the TAMS, 3\,\% at the base of the RGB.

We calculated another evolutionary sequence with the RM12 magnetic braking law. With a coefficient $K_{RM}$ calibrated so as to yield the solar 
rotation rate in a solar model, our $1.3 M_{\odot}$ is much more efficiently slowed down compared to the evolution computed with K88. The central rotation rate,
however, changes by at most 2\,\%.
 
\subsubsection{Stability of the rotation profile}

The changes on the surface rotation rate caused by the mechanisms described above do not change the central
rotation rate significantly because the extraction of angular momentum from the core is not efficient enough. 
The core and the surface are not sufficiently coupled.
One potential way of coupling the core to the surface is through instabilities that might
 arise because of rotation gradient becoming very steep 
(as seen in Fig. \ref{fig:rotgiant}). 
The Rayleigh criterion requires that $r^2 \Omega$ increases outwards.
For our $1.3 M_{\odot}$ model, the rotation profile becomes unstable according to the Rayleigh criterion in two places, at the shell source and
just below the convective envelope. However, the stabilizing effect of the density stratification overcomes this instability, as expressed by the
Solberg-Hoiland criterion:
\begin{equation}
 N^2 = N^2_T + N^2_{\mu} + N^2_{\Omega} > 0,
\end{equation}
where $N^2_{\Omega}$ is the Rayleigh frequency in a rotating medium, given at the equator by
\begin{equation}
N^2_{\Omega} = \frac{1}{r^3} \frac{\d }{\d r} \left(\Omega^2 r^4 \right).
\end{equation}

  \begin{figure}[t]
   \centering
   \includegraphics[width=9cm]{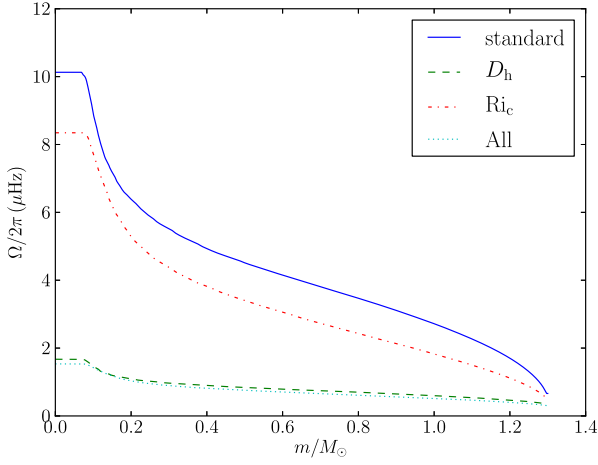}
   \includegraphics[width=9cm]{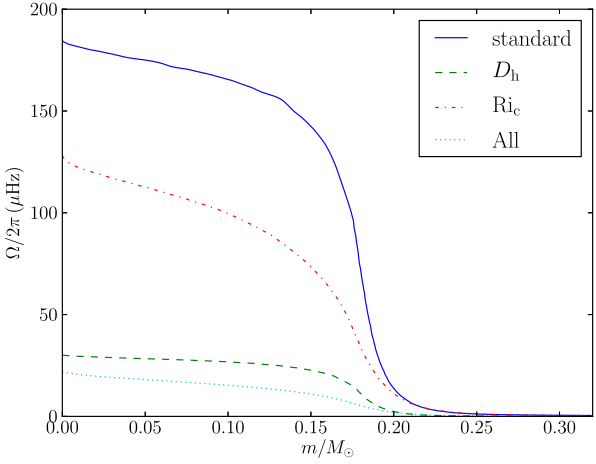}
 \caption{ {\it Top:} Rotation profiles for $1.3 M_{\odot}$ models at 
 the end of the MS calculated assuming standard viscosity coefficients (standard), 
a vertical turbulent viscosity $D\ind V$ computed with ${\rm Ri}_c=1$ (${\rm Ri}_c$), 
a horizontal viscosity coefficient 100 times the standard value ($D\ind h$), 
and all of the above (All). {\it Bottom:} 
The same for models at the base of the RGB when $R_{\star} = 3.73 R_{\odot}$  ($\Delta \nu =21.3 \mu Hz$).}
    \label{fig:visc}
   \end{figure}

According to the Solberg-Hoiland criterion, the rotation profiles we obtain are stable throughout the evolution. However, the Goldreich-Schubert-Fricke
 instability  \citep[GSF, see][]{1967ApJ...150..571G,1968ZA.....68..317F} may occur in stellar
conditions. \citet{2010A&A...519A..16H} have studied its effects on pre-supernova models and find that if $N^2_{\Omega} < 0$ the GSF instability is always present
 regardless of the $\mu-$ and $T-$gradients. We implemented their prescription in CESTAM. Where $N^2_{\Omega} < 0$, turbulent transport by the GSF instability 
operates with a viscosity coefficient given by the solution of their Eq. (20),
\begin{eqnarray}
 \left(N^2_T + N^2_{\mu} + N^2_{\Omega}\right) x^2 + \nonumber \\
 \left[N^2_T \Dh + N^2_{\mu} \left(K + \Dh \right) + N^2_{\Omega} \left(K + 2\Dh \right) \right]x\nonumber \\ + N^2_{\Omega} \left(\Dh K + \Dh^2 \right) = 0,
\end{eqnarray}
where $K$ is the thermal diffusivity given by Eq. (\ref{eq:krad}). The viscosity coefficient associated to the GSF instability is
then given by $D\ind{GSF} = 2x$.

   \begin{figure}[t]
   \centering
   \includegraphics[width=9cm]{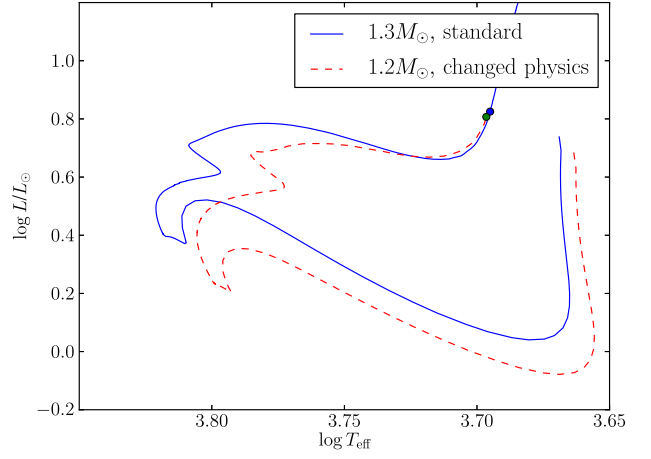}
   \includegraphics[width=9cm]{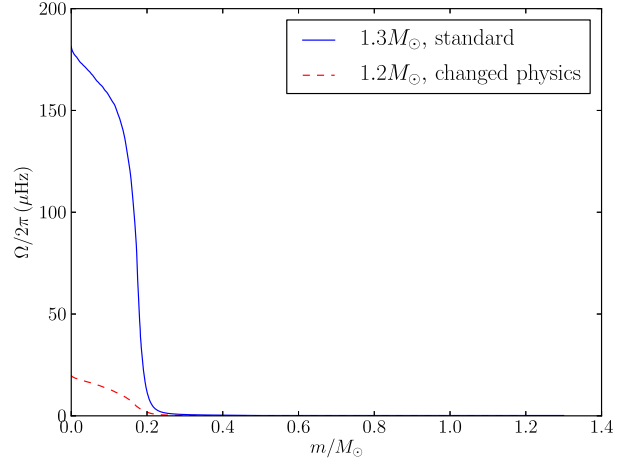}
 \caption{ {\it Top:} HR diagram showing the evolutionary track of a $1.2M_{\odot}$ model computed with
 an overshoot of $0.1 H_P$, a vertical turbulent viscosity $D\ind V$ computed with ${\rm Ri}\ind c=1$, a $D\ind h$ increased by a factor $10^2$ (red line, dashed), 
and a $1.3M_{\odot}$ model
computed without overshooting and with the ``standard'' viscosities (blue line, full). Dots indicate the location of models with  $\Delta \nu=23.3 \, \mu{\rm Hz}$.
{\it Bottom:} Rotation profiles for the models above.
   }
\label{omegatout}
   \end{figure}
  \begin{figure}[h]
   \centering
   \includegraphics[width=8cm]{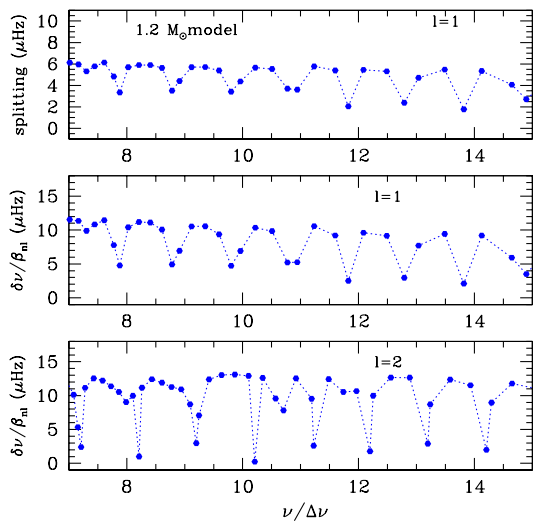}
 \caption{ {\it Top:} Rotational splittings $\delta \nu_{n\ell}$ 
 for $\ell=1$ modes as a function of the normalized frequency 
 $\nu/\Delta \nu$ associated with the rotation profile shown in Fig. \ref{omegatout}
 for the $1.2M_{\odot}$ model.
{\it Middle:} Corresponding $\delta \nu/\beta_{n\ell}$ 
 for $\ell=1$ modes as a function of the normalized frequency 
 $\nu/\Delta \nu$.
  {\it Bottom:} The same for $\ell=2$ modes. }
    \label{split808}
   \end{figure}

We found that $N^2_{\Omega} < 0$ in two regions from the end of the MS, just below the border of the outer CZ and at the shell source. 
At the shell source, the $\Omega-$gradient is high because of the contraction of the layers left behind by the movement of the shell source, while the
descending CZ lowers $\Omega$ just above its border. 
In our models, the GSF turbulent viscosity is on the order of $D\ind V$ in the unstable regions.
The turbulent transport by the GSF instability reduces the $\Omega-$gradient in the unstable region at the shell source, reducing the central rotation
 rate by 3\,\% in models at the base of the RGB. The turbulent transport by the GSF instability does not have time to change the  $\Omega-$gradient below
the CZ before the convective instability sweeps over this region.

\subsubsection{Uncertainties on the turbulent viscosity coefficients}

To extract more angular momentum, we can either increase the meridional circulation or increase the vertical turbulent viscosity.
To increase the meridional circulation, one option is to decrease the inhibiting effect of $\Lambda$ 
(see the expression for the $\mu$-currents, Eq. \ref{eq:emu}). 
According to Eq. (\ref{eq:lambda}), an increase in $\Dh$ will lead to a decrease in $\Lambda$. 
The prescription for the 
horizontal coefficient of turbulence $\Dh$ is still open to discussion. Prescriptions derived by 
\citet{1992A&A...265..115Z}, \citet{2003A&A...399..263M}, and \citet{2004A&A...425..243M} can differ by as
 much as two orders of magnitude. 
We adopted, as a test, a value of $\Dh$ that is $10^2$ higher 
than that previously used (Eq. \ref{eq:dh}). 

The value of the critical Richardson number in Eq. (\ref{eq:dv}) is usually assumed to lie between 1/4 and 1/6.
\citet{2001MNRAS.320...73B} and \citet{2002A&A...384.1119C} indicate, however, that we can
expect instability for ${\rm Ri} \simeq 1$. 
To increase the efficiency of the vertical turbulent viscosity in transporting angular momentum from the core, we 
computed an evolutionary sequence using  ${\rm Ri}\ind c = 1$ to test the effect of 
the uncertainties. Figure \ref{fig:visc} shows profiles of the rotation rate computed with different assumptions of the diffusion coefficients,
at the TAMS and at the base of the RGB when $R_{\star} = 3.73 R_{\odot}$. Models have the same mass and radius, hence the same $\Delta \nu$.
As expected,
with an increased value of ${\rm Ri}_c$ from 1/6 to 1, the central rotation rate decreases by 18\% at the TAMS and by 30\% at the base of
the RGB. With a $D\ind h$ increased by two orders of magnitude (and ${\rm Ri}\ind c=1/6$), the central rotation rate decreases by 84\%
at the TAMS and the base of the RGB. When both changes are included (an increase of ${\rm Ri}\ind c$ and $\Dh$ as above), the central rotation rate 
is decreased by an order of magnitude, while we need 
a decrease of two orders of magnitude to reproduce the observed
rotational splittings.

\subsubsection{A slowly rotating red-giant model}

We eventually computed an evolutionary sequence where we combined  all effects 
causing a decrease in the rotation rate in the central region;
an overshoot of $0.1 \Hp$ and 
a vertical turbulent viscosity $\Dv$ computed with ${\rm Ri}\ind c=1$, 
together with $\Dh$ increased by a factor $10^2$. This model was computed with $M_{\star}=1.2M_{\odot}$ 
because during the RGB the evolutionary track on the HRD of a $1.2M_{\odot}$ 
computed with all these effects lies close to
the evolutionary track of the $1.3M_{\odot}$ models used as comparison.
The resulting  rotation profile for models at the base
of the RGB  with same $\Delta \nu$ is significantly decreased in the central region 
with the central value $ \Omega\ind c =21.5 \mu$ Hz   and 
is displayed in Fig. \ref{omegatout}. 

Calculations of  linear rotational splittings   shows   that the maximum 
splittings amounts to 5-6 $\mu$ Hz for model \#20 at the base of the
RGB  (Fig. \ref{split808}). This corresponds to a  maximum value for
$\delta \nu /\beta_{nl}$ of  $12 \mu$Hz.
The rotation gradient and the value of the central rotation 
 are  now too low for nonperturbative calculations to give rise to 
  significant corrections to  the values of the linear  rotational
 splittings.  Such rotation splittings lead to a core rotation rate that is
  closer to the observations but still too high by almost one order of
magnitude compared with
   what is seen in recent observations 
   \citep{2012Natur.481...55B, 2012ApJ...756...19D, 2012arXiv1209.3336M}.

{\it The seismology of red-giant stars emphasizes that an additional  mechanism is needed to achieve a slower core
   rotation of red-giant
  stars.}

\section{Conclusions and perspectives}\label{sec:conclusions}%

We have computed stellar evolution models of low-mass stars, 
 taking the transport of angular momentum by turbulent viscosity and 
meridional circulation in radiative zones into account. 
As found by \citet{2006A&A...453..261P} and \citet{2010A&A...509A..72E}, the core 
accelerates very rapidly in our models during the subgiant and giant phases.
 We showed that the  linear approximation for computing rotational splittings 
remains nevertheless valid for  stochastically excited solar-type modes, for 
all evolutionary stages until the RGB.
 
For red-giant stars, interpretations of recent \Kepler observations lead to 
a ratio between the core and surface rotation rates of five to ten,  far lower 
than in our models  where the ratio between the core and surface rotation rates approaches~$10^3$. 
Similar conclusions were reached by \cite{2012A&A...544L...4E}, which was published during the
 submission process of the present paper. 
Nonperturbative calculations lead to
complex frequency spectra, with in particular nonsymmetric multiplets
 \citep[see for instance without Coriolis force][with both centrifugal and Coriolis force]{2006A&A...455..607L,2006A&A...455..621R,Ouazzani2012}.
 Observations show that several red-giant
stars do indeed have such complex spectra. Their core rotation is fast enough to have entangled rotational splittings
and mixed-mode spacings. In such cases, rotation may have to be studied with nonperturbative methods. 
On the other hand, other red-giant
stars  \citep{2012Natur.481...55B} show frequency spectra where 
the rotation splittings are easily identified as  symmetric
 patterns around axisymmetric modes. The values of the corresponding splittings 
 are quite low, close to or smaller than   $1 \mu$Hz, 
 and using the linear approximation  to derive the  rotation splittings from stellar models 
 is fully justified.  
 
 We have computed the evolution of a stellar model including  rotationally induced transport 
  taking uncertainties in the values of the  parameters entering such a description into account. 
   We achieved a decrease in the central rotation rate of a red-giant star by an order of magnitude at most, mostly
due to an increase in $\Dh$ by two orders of magnitude. The modification of
  the parameters were chosen so as to enhance the transport of angular momentum so that 
  the core rotation is slowed down  as much as possible.  The models in the red-giant phase were then
   slowed  enough that the calculations of linear rotation splittings are  valid. They were
  found to be one order of magnitude larger still than observed. This indicates  
 that extraction of 
 angular momentum  from the core   in our models is not efficient enough.

We need to decrease it by another 
order of magnitude to reproduce the observations. A similar situation is encountered  in the case of the Sun. Several possibilities 
have been proposed, the main candidates
 being  magnetic fields and internal gravity waves. We showed that such mechanisms
  must operate or keep operating  during the whole subgiant 
and red-giant phase. In this context, 
\citet{2008A&A...482..597T}  show that  transport of angular momentum 
by internal waves generated by the convective envelope 
can  play  an important role from the subgiant phase to the base of the 
 RGB.  They also find that this transport 
  seems to have no major impact on stars  ascending the RGB itself and later on.
 In the RGB phase, another mechanism  must be called for to explain the observations.
This will be investigated further in future work. On the observational side, 
measurements of rotational splittings for subgiant stars are 
crucially needed  to constrain
 the efficiency  for their central layers to slow down.


A standard, validated version of CESTAM will be available for download soon, 
together with  grids of stellar models  computed with
several rotational velocities and adiabatic oscillation frequencies for selected models.

 \begin{acknowledgements}
 JPM acknowledges financial support through a 3-year CDD contract with the CNES. 
 The authors also acknowledge financial
support from the French National Research Agency (ANR) for the project ANR-07-BLAN-0226 SIROCO (SeIsmology, 
ROtation and COnvection with the CoRoT satellite). A.M., acknowledges the support of the Direccion General de Investigacion 
under project ESP2004-03855-C03-01. He also acknowledges a stay of two 
years at the Observatoire de Paris-Meudon.
\end{acknowledgements}

\appendix

\section{Meridional circulation} \label{sec:detail}%

The vertical component of the meridional circulation is given by

\begin{eqnarray}
 U_2 &=& \frac{p}{c_p \rho T g \left[\nabla_{\rm ad} - \nabla + (\varphi/\delta) \nabla_{\mu} \right]} 
\left[\frac{L_r}{M^{\star}} \left(E_{\Omega} + E_{\mu} \right) + \frac{c_p T}{\delta} \frac{\d \Theta}{\d t} \right], \nonumber \\
&& \label{eq:u2}
\end{eqnarray}
where $c_p, T, g, \nabla $, and $\nabla_{\rm ad}$ have their usual meanings \citep[as defined in][]{Kippenhahn/Weigert:1991}.
In Eq. (\ref{eq:u2}) above, $M^{\star}$ is the reduced mass,
\begin{equation}
 M^{\star} = m\left(1 - \frac{\Omega^2}{2 \pi G \rho_m} \right),
\end{equation}
where $m$ and $\rho_m$ are the mass and mean density inside an isobar, respectively, and
$E_{\Omega}$ and $E_{\mu}$ denote 
the so-called $\Omega$- and $\mu$-currents 
\citep[following][]{2003A&A...399..603P}. The quantities $\varphi$ and $\delta$ are obtained from the equation of state, and
 are defined as in \citet{Kippenhahn/Weigert:1991},
\begin{equation}
 \delta = -\left(\pderiv{\ln \rho}{\ln T} \right)_{P,\mu} \, ; \, \varphi = \left(\pderiv{\ln \rho}{\ln \mu} \right)_{P,T}.
\end{equation}

The $\Omega$- and $\mu$-currents are given by
\begin{eqnarray}
E_{\Omega} &=& 2 \left(1 - \frac{\Omega^2}{2 \pi G \rho} - \frac{\bar{\varepsilon} + \bar{\varepsilon}_g}{\varepsilon_m} \right) \frac{\tilde{g}}
{\bar{g}} - \nonumber \\
&& - \frac{\rho_m}{\rho} \left\{ \frac{r}{3} \frac{\partial}{\partial r} \left[H_T \frac{\partial}{\partial r}\left(\frac{\Theta}{\delta} \right) - 
\chi_T \frac{\Theta}{\delta} + \left(1 - \frac{1}{\delta} \right) \Theta \right] - \right . \nonumber \\
&& \left . - \frac{2 H_T}{r}\left(1 + \frac{\Dh}{K} \right)\frac{\Theta}{\delta} + \frac 2 3 \Theta \right\} -\nonumber \\
&&  - \frac{\bar{\varepsilon} + \bar{\varepsilon}_g}{\varepsilon_m} \left[H_T \frac{\partial}{\partial r} \left( \frac{\Theta}{\delta} \right) + 
\left(f_{\varepsilon} \varepsilon_T - \chi_T  \right) \frac{\Theta}{\delta} + \right. \nonumber \\
&& \left. + \left(2 - f_{\varepsilon} - \frac{1}{\delta} \right)\Theta \right] \label{eq:eomega} \\
E_{\mu} &=& \frac{\rho_m}{\rho} \left\{\frac{r}{3} \frac{\partial}{\partial r} \left[H_T \frac{\partial}{\partial r} \left(\frac{\varphi}{\delta} 
\Lambda \right) - \left(\chi_{\mu} + \frac{\varphi}{\delta} \chi_T +  \frac{\varphi}{\delta} \right)\Lambda \right] - \right . \nonumber \\
&& - \left . \frac{2 H_T}{r} \frac{\varphi}{\delta}\Lambda \right\} + \frac{\bar{\varepsilon} + \bar{\varepsilon}_g}{\varepsilon_m} 
\left\{H_T \frac{\partial}{\partial r}\left(\frac{\varphi}{\delta} \Lambda \right) + \right .  \nonumber \\
&& \left . \left[f_{\varepsilon} \left(\varepsilon_{\mu} + \frac{\varphi}{\delta} \varepsilon_{T} \right) - \chi_{\mu} - 
\frac{\varphi}{\delta}\left(\chi_T + 1 \right)\right]\Lambda \right\} \label{eq:emu}.
\end{eqnarray}

In Eqs. (\ref{eq:eomega}) and (\ref{eq:emu}) we used the following quantities:
$\bar{\varepsilon}$ and $\bar{\varepsilon}_g$ are the nuclear and gravitational 
energy generation rates, $\varepsilon_m=L/m$ the mean energy production inside a sphere of radius $r$, 
and $\rho_m=3 m/(4\pi r^3)$ the mean density. The thermal conductivity $\chi$ is given by
\begin{equation}
 \chi = \frac{4 a c T^3}{3 \kappa \rho},
\end{equation}
 and the thermal diffusivity is 
\begin{equation}
K = \frac{\chi}{\rho c_P} = \frac{4 a c T^3}{3 \kappa \rho^2 c_P}. \label{eq:krad}
\end{equation}

We also used
\begin{equation}
 H_T = -\tderiv{r}{\ln T},
\end{equation}
the temperature scale height; 
\begin{equation}
 \chi_{T} = \left(\pderiv{\ln \chi}{\ln T} \right)_{P,\mu}\,{\rm and}\,\,\chi_{\mu} = \left(\pderiv{\ln \chi}{\ln \mu} \right)_{P,T},
\end{equation}
the logarithmic derivatives of $\chi$;
\begin{equation}
 \varepsilon_{T} = \left(\pderiv{\ln \varepsilon}{\ln T} \right)_{P,\mu}\,{\rm and}\,\,\varepsilon_{\mu} = \left(\pderiv{\ln \varepsilon}{\ln \mu} \right)_{P,T},
\end{equation}
the logarithmic derivatives of $\varepsilon$; and
\begin{equation}
 f_{\varepsilon} = \frac{\bar{\varepsilon}}{\bar{\varepsilon}+\bar{\varepsilon}_g}.
\end{equation} 

\section{Turbulent coefficients of diffusivity} \label{sec:visc}%

The coefficient $D_{\rm eff}$ is given by \citep{1992A&A...253..173C}
\begin{equation}
 D_{\rm eff} = \frac{\left(r U_2 \right)^2}{30 \Dh}.
\end{equation}

We use the prescription of \citet{2004A&A...425..243M} for the horizontal component of turbulent diffusivity $D_h$:
\begin{equation}
 \Dh^2 = \frac{\beta}{10} r^3 \Omega \left|2 V_2 - \alpha U_2 \right|, \label{eq:dh}
\end{equation}
where 
\begin{equation}
 \alpha = \frac 1 2 \pderiv{\ln r^2 \Omega}{\ln r}.
\end{equation}
We take the coefficient $\beta = 1.5 \times 10^{-5}$ \citep{1999A&A...347..734R}.

For the vertical component $\Dv$, we use the prescription of
 \citet{1997A&A...317..749T},
\begin{equation}
 \Dv = \frac{{\rm Ri}\ind c \left(K + \Dh \right) r^2}{N^2_T + N^2_{\mu} \left(1 + K/\Dh \right)} \left(\pderiv{\Omega}{r} \right)^2, \label{eq:dv}
\end{equation}
where ${\rm Ri}\ind c = 1/6$ is the critical Richardson
 number, $K$ the thermal diffusivity, $N_T$ and $N_{\mu}$ 
 the chemical and thermal parts of the Brunt-V\H{a}is\H{a}l\H{a}
  frequency, $N^2 = N^2_T + N^2_{\mu}$:
\begin{equation}
  N^2_T = \frac{g \delta}{H_P} \left(\nabla_{\rm ad} - \nabla \right); \,\,\, N^2_{\mu} = \frac{g \varphi}{H_P}\nabla_{\mu}.
\end{equation}

\bibliography{biblio}
\bibliographystyle{aa} 

\end{document}